\documentclass[11pt]{article}
\usepackage{epsfig}
\usepackage{feynmf}
\usepackage{amssymb}
\begin{document}
\newcommand{\iint}{\int \! \! \int}
\newcommand{\NP}[1]{Nucl.\ Phys.\ {\bf #1}}
\newcommand{\PL}[1]{Phys.\ Lett.\ {\bf #1}}
\newcommand{\NC}[1]{Nuovo Cimento {\bf #1}}
\newcommand{\CMP}[1]{Comm.\ Math.\ Phys.\ {\bf #1}}
\newcommand{\PR}[1]{Phys.\ Rev.\ {\bf #1}}
\newcommand{\PRL}[1]{Phys.\ Rev.\ Lett.\ {\bf #1}}
\newcommand{\ZPH}[1]{Z.\ Phys.\  {\bf #1}}
\newcommand{\AP}[1]{Ann.\ Phys.\ {\bf #1}}

\title{\bf \Large A full Next to Leading Order study of \\
direct photon pair production in hadronic collisions}
\author{T.~Binoth, J.Ph.~Guillet, E.~Pilon, M.~Werlen}
\date{Laboratoire d'Annecy-le-Vieux de Physique Th\'eorique LAPTH
\footnote{UMR 5108 du
CNRS, associ\'ee \`a l'Universit\'e de Savoie.}\\
B.P. 110, F-74941 Annecy-le-Vieux Cedex, France} 
\maketitle

\begin{abstract}
We discuss the production of photon pairs in hadronic collisions, from 
fixed target to LHC energies. The study which follows is based on a 
QCD calculation at full next-to-leading order accuracy, including single
and double fragmentation contributions, and implemented in the form of a
general purpose computer program of ``partonic event generator" type.
To illustrate the possibilities of this code, we present the comparison
with observables measured by the WA70 and D0 collaborations,
and some predictions for the irreducible background to the search of
Higgs bosons at LHC in the channel $h \rightarrow \gamma \gamma$. We also
discuss theoretical scale uncertainties for these predictions, and examine
several infrared sensitive situations which deserve further study.

\end{abstract}

\vskip 4 truecm
\rightline{LAPTH-760/99}
\thispagestyle{empty}
\clearpage

\setcounter{page}{1}

\section{Introduction}

The production of pairs of direct photons\footnote{The word
``direct" means here that these photons do not result from the decay of
$\pi^{0}$, $\eta$, $\omega$ at large transverse momentum. Direct photons may be
produced according to two possible mechanisms: either they take part
directly to the hard subprocess, or they result from the fragmentation
of partons themselves produced at high transverse momentum in the
subprocess; see sect. 2.} with large invariant mass is the so called
irreducible background for the search of the Higgs boson in the two
photon decay channel in the intermediate mass range $80$ GeV $\leq m_{h} \leq 
140$ GeV at the forthcoming LHC. This background is huge and requires to be
understood and quantitatively evaluated.\\

Beside this important motivation, this process deserves interest by its own. 
The production of such pairs of photons has been experimentally studied in a 
large domain of energies, from fixed targets \cite{wa70cross,wa70correl,e706}
to colliders \cite{ua2,cdf1,d0}. A wide variety of observables has been
measured,  such as distributions of invariant mass, azimuthal angle and 
transverse momentum of the pairs of photons, inclusive transverse momentum 
distributions of each photon, which offer the opportunity to test our 
understanding of this process.\\

The aim of this article is to present a study of diphoton hadroproduction
based on a computer code of partonic event generator type. In this code, we
account for all contributing processes consistently at next-to-leading order
(NLO) accuracy, together with the so called box contribution $gg \rightarrow
\gamma \gamma$. This code is flexible enough to accommodate various kinematic or
calorimetric cuts. Especially, it allows to compute cross sections for both
inclusive and isolated direct photon pairs, for any infrared and collinear safe
isolation criterion which can be implemented at the partonic level. This
article is organized according to the following outline. In section 2, we
remind the basic theoretical ingredients, and present the method used to build
the computer code developed for this study. Section 3 is dedicated to the
phenomenology of photon pair production. We start with a comparison with fixed 
target and collider experiments. We then provide some predictions for LHC, 
together with a discussion of theoretical scale uncertainties. The theoretical
discussion about the present day limitations of our code is continued in
section 4. There we mention various infrared  sensitive situations, which would
deserve some more care, and for which the resummation of multiple soft gluon
effects would be required, in order to improve the ability of our code to
account for such observables. Section 5 gathers our conclusions and
perspectives.

\begin{fmffile}{samplepics}

\section{Theoretical content and presentation of the method}

Let us first remind briefly the theoretical level of accuracy 
and limitations of works prior to the present one, in order to assess 
the improvements which we introduce. Then we present the method which 
we used to build our computer code {\it DIPHOX}.

\subsection{Theoretical content}\label{theorcont}

The theoretical understanding of this process relies on NLO calculations, 
initiated in \cite{abfs}. The leading order contribution to diphoton
reactions is given  by the Born level process  $q \bar{q} \rightarrow \gamma
\gamma$ see for instance Diagram a. The computation of NLO contributions to it
yields ${\cal O}(\alpha_s)$  corrections coming from the subprocesses  $q
\bar{q} \rightarrow \gamma \gamma g$, $g q$ (or $\bar{q}$) $\rightarrow
\gamma \gamma q$ (or $\bar{q}$) and corresponding virtual corrections, see for
example Diagrams b and c. \\

\[
\parbox[c][60mm][c]{40mm}{\begin{fmfgraph*}(40,40)
  \fmfleft{H1,H2}
  \fmfright{Hf1,ph1,ph2,Hf2}
  \fmf{dbl_plain_arrow,tension=2,lab=$p$}{H1,VP1}
  \fmf{dbl_plain_arrow,tension=2,lab=$p$}{H2,VP2}
  \fmf{plain}{VP1,Hf1}
  \fmf{plain}{VP2,Hf2}
  \fmfblob{.16w}{VP1,VP2}
  \fmf{fermion}{VP1,v1}
  \fmf{fermion}{v1,v2}
  \fmf{fermion}{v2,VP2}
  \fmf{photon}{v1,ph1}
  \fmf{photon}{v2,ph2}
  \fmfv{label=Diagram a,label.angle=-90.,label.dist=0.3w}{VP1}
  \fmffreeze
  \fmfi{plain}{vpath (__VP1,__Hf1) shifted (thick*(0,2))}
  \fmfi{plain}{vpath (__VP1,__Hf1) shifted (thick*(0,-2))}
  \fmfi{plain}{vpath (__VP2,__Hf2) shifted (thick*(0,2))}
  \fmfi{plain}{vpath (__VP2,__Hf2) shifted (thick*(0,-2))}
\end{fmfgraph*}}
\qquad + \cdots + \qquad
\parbox[c][60mm][c]{40mm}{\begin{fmfgraph*}(40,40)
  \fmfleft{H1,H2}
  \fmfright{Hf1,ph1,ph2,Hf2}
  \fmf{dbl_plain_arrow,tension=2,lab=$p$}{H1,VP1}
  \fmf{dbl_plain_arrow,tension=2,lab=$p$}{H2,VP2}
  \fmf{plain}{VP1,Hf1}
  \fmf{plain}{VP2,Hf2}
  \fmfblob{.16w}{VP1,VP2}
  \fmf{fermion}{VP1,vi1,v1}
  \fmf{fermion}{v1,v2}
  \fmf{fermion}{v2,vi2,VP2}
  \fmf{photon}{v1,ph1}
  \fmf{photon}{v2,ph2}
  \fmfv{label=Diagram b,label.angle=-90.,label.dist=0.3w}{VP1}
  \fmffreeze
  \fmf{gluon}{vi1,vi2}
  \fmfi{plain}{vpath (__VP1,__Hf1) shifted (thick*(0,2))}
  \fmfi{plain}{vpath (__VP1,__Hf1) shifted (thick*(0,-2))}
  \fmfi{plain}{vpath (__VP2,__Hf2) shifted (thick*(0,2))}
  \fmfi{plain}{vpath (__VP2,__Hf2) shifted (thick*(0,-2))}
\end{fmfgraph*}}
\qquad + \cdots 
\]
\[
+ \qquad
\parbox[c][60mm][c]{40mm}{\begin{fmfgraph*}(40,40)
  \fmfleft{H1,H2}
  \fmfright{Hf1,qq,ph1,ph2,Hf2}
  \fmf{dbl_plain_arrow,tension=2,lab=$p$}{H1,VP1}
  \fmf{dbl_plain_arrow,tension=2,lab=$p$}{H2,VP2}
  \fmf{plain}{VP1,Hf1}
  \fmf{plain}{VP2,Hf2}
  \fmfblob{.16w}{VP1,VP2}
  \fmf{gluon,tension=1.5}{VP2,v1}
  \fmf{fermion}{VP1,v1,v2,vi,qq}
  \fmf{photon}{v2,ph2}
  \fmfv{label=Diagram c,label.angle=-90.,label.dist=0.3w}{VP1}
  \fmffreeze
  \fmf{photon}{vi,ph1}
  \fmfi{plain}{vpath (__VP1,__Hf1) shifted (thick*(0,2))}
  \fmfi{plain}{vpath (__VP1,__Hf1) shifted (thick*(0,-2))}
  \fmfi{plain}{vpath (__VP2,__Hf2) shifted (thick*(0,2))}
  \fmfi{plain}{vpath (__VP2,__Hf2) shifted (thick*(0,-2))}
\end{fmfgraph*}}
\qquad + \cdots
\]
\vspace{0.5cm}

Yet it also yields the leading order contribution of single fragmentation type
(sometimes called ``Bremsstrahlung contribution"), in which one of the photons
comes from the collinear fragmentation of a hard parton produced in the short
distance subprocess, see for example Diagram d. From a physical point of view
such a photon is most probably accompanied by hadrons. From a technical point
of view, a final state quark-photon collinear singularity appears in the
calculation of the contribution from the subprocess $g q \rightarrow \gamma 
\gamma q$. At higher orders, final state multiple collinear singularities
appear in any subprocess where a high $p_{T}$ parton (quark or gluon)
undergoes a cascade of successive collinear splittings ending up with a
quark-photon splitting. These singularities are factorized to all orders in 
$\alpha_s$ according to the factorization property, and absorbed into quark and
gluon fragmentation functions to a photon $D_{\gamma/q \; or \;
g}(z,M_{f}^{2})$ defined in some arbitrary fragmentation scheme, at some
arbitrary fragmentation scale $M_{f}$. When the fragmentation
scale $M_{f}$, chosen of the order of the hard scale of the subprocess, is
large compared to any typical hadronic scale $\sim 1 $~GeV, these functions
behave roughly as $\alpha/\alpha_s(M_{f}^{2})$. Then a power counting argument
tells that these contributions are asymptotically of the same order in
$\alpha_s$ as the Born term $q \bar{q} \rightarrow \gamma \gamma$. What is
more, given the high gluon luminosity at LHC, the $g q$ (or $\bar{q}$)
initiated contribution involving one photon from fragmentation even dominates
the inclusive production rate in the invariant mass range $80$ GeV $ \leq
m_{\gamma \gamma} \leq 140$  GeV. A consistent treatment of diphoton
production at NLO thus requires that ${\cal O}(\alpha_s)$ corrections to these
contributions be calculated also, see for example Diagrams e and f. They have
not been incorporated in \cite{abfs,owens,yuan}, and we compute them in the
present work. \\

\[
\parbox[c][60mm][c]{40mm}{\begin{fmfgraph*}(40,40)
  \fmfleft{H1,H2}
  \fmfright{Hf1,ph1,ph2,Hf2}
  \fmf{dbl_plain_arrow,tension=2,lab=$p$}{H1,VP1}
  \fmf{dbl_plain_arrow,tension=2,lab=$p$}{H2,VP2}
  \fmf{plain}{VP1,Hf1}
  \fmf{plain}{VP2,Hf2}
  \fmfblob{.16w}{VP1,VP2}
  \fmf{gluon,tension=1.5}{VP2,v1}
  \fmf{fermion}{VP1,v1}
  \fmf{fermion,tension=2}{v1,v2}
  \fmf{fermion}{v2,vi}
  \fmf{photon}{vi,ph1}
  \fmf{photon}{v2,ph2}
  \fmfblob{.10w}{vi}
  \fmfv{lab=$D_{\gamma/q}$,lab.dist=.1w}{vi}
  \fmfv{label=Diagram d,label.angle=-90.,label.dist=0.3w}{VP1}
  \fmffreeze
  \fmfi{plain}{vpath (__VP1,__Hf1) shifted (thick*(0,2))}
  \fmfi{plain}{vpath (__VP1,__Hf1) shifted (thick*(0,-2))}
  \fmfi{plain}{vpath (__VP2,__Hf2) shifted (thick*(0,2))}
  \fmfi{plain}{vpath (__VP2,__Hf2) shifted (thick*(0,-2))}
\end{fmfgraph*}}
\qquad + \cdots + \qquad
\parbox[c][60mm][c]{40mm}{\begin{fmfgraph*}(40,40)
  \fmfleft{H1,H2}
  \fmfright{Hf1,ph1,ph2,Hf2}
  \fmf{dbl_plain_arrow,tension=2,lab=$p$}{H1,VP1}
  \fmf{dbl_plain_arrow,tension=2,lab=$p$}{H2,VP2}
  \fmf{plain}{VP1,Hf1}
  \fmf{plain}{VP2,Hf2}
  \fmfblob{.16w}{VP1,VP2}
  \fmf{gluon,tension=1.5}{VP2,v1}
  \fmf{fermion}{VP1,vi1}
  \fmf{fermion}{vi1,v1}
  \fmf{fermion}{v1,v2}
  \fmf{fermion,tension=.1}{v2,vi2}
  \fmf{fermion}{vi2,vi}
  \fmf{photon}{vi,ph1}
  \fmf{photon}{v2,ph2}
  \fmfblob{.10w}{vi}
  \fmfv{lab=$D_{\gamma/q}$,lab.dist=.1w,lab.ang=-90}{vi}
  \fmf{gluon,tension=.5}{vi1,vi2}
  \fmfv{label=Diagram e,label.angle=-90.,label.dist=0.3w}{VP1}
  \fmffreeze
  \fmf{gluon}{vi1,vi2}
  \fmfi{plain}{vpath (__VP1,__Hf1) shifted (thick*(0,2))}
  \fmfi{plain}{vpath (__VP1,__Hf1) shifted (thick*(0,-2))}
  \fmfi{plain}{vpath (__VP2,__Hf2) shifted (thick*(0,2))}
  \fmfi{plain}{vpath (__VP2,__Hf2) shifted (thick*(0,-2))}
\end{fmfgraph*}}
\qquad + \cdots 
\]
\[
+ \qquad
\parbox[c][60mm][c]{40mm}{\begin{fmfgraph*}(40,40)
  \fmfleft{H1,H2}
  \fmfright{Hf1,ph1,ph2,qq,Hf2}
  \fmf{dbl_plain_arrow,tension=2,lab=$p$}{H1,VP1}
  \fmf{dbl_plain_arrow,tension=2,lab=$p$}{H2,VP2}
  \fmf{plain}{VP1,Hf1}
  \fmf{plain}{VP2,Hf2}
  \fmfblob{.16w}{VP1,VP2}
  \fmf{gluon}{VP2,v1}
  \fmf{gluon}{VP1,v2}
  \fmf{fermion}{qq,vi1,v1,v2,vi}
  \fmf{photon}{vi,ph1}
  \fmfblob{.10w}{vi}
  \fmfv{label=Diagram f,label.angle=-90.,label.dist=0.3w}{VP1}
  \fmffreeze
  \fmf{photon}{vi1,ph2}
  \fmfv{lab=$D_{\gamma/q}$,lab.dist=.1w,lab.ang=-90}{vi}
  \fmfi{plain}{vpath (__VP1,__Hf1) shifted (thick*(0,2))}
  \fmfi{plain}{vpath (__VP1,__Hf1) shifted (thick*(0,-2))}
  \fmfi{plain}{vpath (__VP2,__Hf2) shifted (thick*(0,2))}
  \fmfi{plain}{vpath (__VP2,__Hf2) shifted (thick*(0,-2))}
\end{fmfgraph*}}
\qquad + \cdots
\]
\vspace{0.5cm}

The calculation of these corrections in their turn yields the leading order 
contribution of yet another mechanism, of double fragmentation type, see for
example Diagram g.  In the latter case, both photons result from the collinear
fragmentation of a hard parton. In order to present a study of  consistent NLO
accuracy, NLO corrections to this double fragmentation  contribution, see for
example Diagrams h and i, have to be calculated accordingly. This is also done
in the present article. \\

\[
\parbox[c][60mm][c]{40mm}{\begin{fmfgraph*}(40,40)
  \fmfleft{H1,H2}
  \fmfright{Hf1,ph1,ph2,Hf2}
  \fmf{dbl_plain_arrow,tension=2,lab=$p$}{H1,VP1}
  \fmf{dbl_plain_arrow,tension=2,lab=$p$}{H2,VP2}
  \fmf{plain}{VP1,Hf1}
  \fmf{plain}{VP2,Hf2}
  \fmfblob{.16w}{VP1,VP2}
  \fmf{gluon}{VP2,v1}
  \fmf{gluon}{VP1,v2}
  \fmf{fermion}{vi1,v1,v2,vi2}
  \fmf{photon}{vi2,ph1}
  \fmfblob{.10w}{vi1,vi2}
  \fmf{photon}{vi1,ph2}
  \fmfv{lab=$D_{\gamma/q}$,lab.dist=.1w,lab.ang=90}{vi1}
  \fmfv{lab=$D_{\gamma/q}$,lab.dist=.1w,lab.ang=-90}{vi2}
  \fmfv{label=Diagram g,label.angle=-90.,label.dist=0.3w}{VP1}
  \fmffreeze
  \fmfi{plain}{vpath (__VP1,__Hf1) shifted (thick*(0,2))}
  \fmfi{plain}{vpath (__VP1,__Hf1) shifted (thick*(0,-2))}
  \fmfi{plain}{vpath (__VP2,__Hf2) shifted (thick*(0,2))}
  \fmfi{plain}{vpath (__VP2,__Hf2) shifted (thick*(0,-2))}
\end{fmfgraph*}}
\qquad + \cdots + \qquad
\parbox[c][60mm][c]{40mm}{\begin{fmfgraph*}(40,40)
  \fmfleft{H1,H2}
  \fmfright{Hf1,ph1,ph2,Hf2}
  \fmf{dbl_plain_arrow,tension=2,lab=$p$}{H1,VP1}
  \fmf{dbl_plain_arrow,tension=2,lab=$p$}{H2,VP2}
  \fmf{plain}{VP1,Hf1}
  \fmf{plain}{VP2,Hf2}
  \fmfblob{.16w}{VP1,VP2}
  \fmf{gluon}{VP2,v1}
  \fmf{gluon}{VP1,v2}
  \fmf{fermion}{vi1,vii1,v1,v2,vii2,vi2}
  \fmf{photon}{vi2,ph1}
  \fmfblob{.10w}{vi1,vi2}
  \fmf{photon}{vi1,ph2}
  \fmfv{lab=$D_{\gamma/q}$,lab.dist=.1w,lab.ang=90}{vi1}
  \fmfv{lab=$D_{\gamma/q}$,lab.dist=.1w,lab.ang=-90}{vi2}
  \fmfv{label=Diagram h,label.angle=-90.,label.dist=0.3w}{VP1}
  \fmffreeze
  \fmf{gluon}{vii1,vii2}
  \fmfi{plain}{vpath (__VP1,__Hf1) shifted (thick*(0,2))}
  \fmfi{plain}{vpath (__VP1,__Hf1) shifted (thick*(0,-2))}
  \fmfi{plain}{vpath (__VP2,__Hf2) shifted (thick*(0,2))}
  \fmfi{plain}{vpath (__VP2,__Hf2) shifted (thick*(0,-2))}
\end{fmfgraph*}}
\qquad + \cdots 
\]
\[
+ \qquad
\parbox[c][60mm][c]{40mm}{\begin{fmfgraph*}(40,40)
  \fmfleft{H1,H2}
  \fmfright{Hf1,ph1,qq,ph2,Hf2}
  \fmf{dbl_plain_arrow,tension=2,lab=$p$}{H1,VP1}
  \fmf{dbl_plain_arrow,tension=2,lab=$p$}{H2,VP2}
  \fmf{plain}{VP1,Hf1}
  \fmf{plain}{VP2,Hf2}
  \fmfblob{.16w}{VP1,VP2}
  \fmf{gluon}{VP2,v2}
  \fmf{gluon}{VP1,v1}
  \fmf{gluon}{v1,vii}
  \fmf{gluon}{vii,v2}
  \fmf{gluon}{v2,vi2}
  \fmf{gluon}{v1,vi1}
  \fmf{photon}{vi2,ph2}
  \fmf{photon}{vi1,ph1}
  \fmfblob{.10w}{vi1,vi2}
  \fmfv{lab=$D_{\gamma/g}$,lab.dist=.1w}{vi1}
  \fmfv{lab=$D_{\gamma/g}$,lab.dist=.1w}{vi2}
  \fmfv{label=Diagram i,label.angle=-90.,label.dist=0.3w}{VP1}
  \fmffreeze
  \fmf{gluon}{vii,qq}
  \fmfi{plain}{vpath (__VP1,__Hf1) shifted (thick*(0,2))}
  \fmfi{plain}{vpath (__VP1,__Hf1) shifted (thick*(0,-2))}
  \fmfi{plain}{vpath (__VP2,__Hf2) shifted (thick*(0,2))}
  \fmfi{plain}{vpath (__VP2,__Hf2) shifted (thick*(0,-2))}
\end{fmfgraph*}}
\qquad + \cdots
\]
\vspace{0.5cm}

We call ``two direct" the contribution given by the Born term plus  the
fraction of the higher order corrections  from which final state collinear
singularities have been subtracted according to the ${\overline{MS}}$
factorization scheme.
We call ``one fragmentation" (``two fragmentation") the  contribution involving
one single fragmentation function (two fragmentation functions) of a parton
into a photon. Let us add one more comment about the splitting into these
three mechanisms. One must keep in mind that this distinction is schematic
and  ambiguous. We remind that it comes technically from the appearance of
final state collinear singularities, which are factorized and absorbed into
fragmentation functions at some arbitrary fragmentation scale\footnote{More
generally, the definition of the fragmentation functions rely on the choice of
a given factorization scheme, e.g. the ${\overline{MS}}$ scheme in this work. 
The fragmentation functions  which we use are presented in \cite{phofrag}.} 
$M_{f}$. Each of the contributions associated with these three mechanisms thus
depends on this arbitrary scale. This dependence on $M_{f}$ cancels
only in the sum of the three, so that this sum only is a physical observable.
More precisely, a calculation of these contributions beyond leading order is
required to obtain a (partial) cancellation of the dependence on $M_{f}$. Indeed
this cancellation starts to occur between the higher order of the ``two direct"
contribution and the leading order of the ``one fragmentation" term, and
similarly between the ``one-" and ``two fragmentation" components respectively.
This is actually one of the first motivations of the present work. Thus, even
though it may be suggestive  to compare the respective sizes and shapes of the
separate contributions  for a given choice of scale, as will be done in \ref{scaledep}, we emphasize that only their sum is meaningful.\\

\[
\parbox[c][60mm][c]{40mm}{\begin{fmfgraph*}(40,40)
  \fmfleft{H1,H2}
  \fmfright{Hf1,ph1,ph2,Hf2}
  \fmf{dbl_plain_arrow,tension=2,lab=$p$}{H1,VP1}
  \fmf{dbl_plain_arrow,tension=2,lab=$p$}{H2,VP2}
  \fmf{plain}{VP1,Hf1}
  \fmf{plain}{VP2,Hf2}
  \fmfblob{.16w}{VP1,VP2}
  \fmf{gluon,tension=1.5}{VP2,v1}
  \fmf{gluon,tension=1.5}{VP1,vi1}
  \fmf{fermion}{vi1,v1}
  \fmf{fermion}{v1,v2}
  \fmf{fermion,tension=0.2}{v2,vi2}
  \fmf{photon,tension=1.5}{vi2,ph1}
  \fmf{photon,tension=1.5}{v2,ph2}
  \fmf{fermion,tension=1}{vi1,vi2}
  \fmfv{label=Diagram j,label.angle=-90.,label.dist=0.3w}{VP1}
  \fmffreeze
  \fmfi{plain}{vpath (__VP1,__Hf1) shifted (thick*(0,2))}
  \fmfi{plain}{vpath (__VP1,__Hf1) shifted (thick*(0,-2))}
  \fmfi{plain}{vpath (__VP2,__Hf2) shifted (thick*(0,2))}
  \fmfi{plain}{vpath (__VP2,__Hf2) shifted (thick*(0,-2))}
\end{fmfgraph*}}
\qquad + \cdots
\]
\vspace{0.5cm}

Beyond this, the ${\cal O}(\alpha_{s}^{2})$ so-called box contribution  $g g
\rightarrow \gamma \gamma$ through a quark loop is also included, see for
example Diagram j. Strictly speaking it is a NNLO contribution from the point of
view of power counting. However in the  range of interest at LHC for the search
of the Higgs boson, the gluon luminosity is  so large compared with the quark
and antiquark one, that it nearly compensates the extra powers of $\alpha_{s}$,
so as to yield a contribution comparable with the Born term. For this reason,
it has been included in previous works, and will be in the present one as
well. We define the ``direct" contribution as the sum ``two direct" + box.\\

Actually one should notice, firstly, that other NNLO gluon-gluon initiated 
processes, such as the collinear finite part of  $g g \rightarrow \bar{q} q
\gamma \gamma$ have been ignored\footnote{The  collinear divergent parts of
these $2 \rightarrow 4$ processes have been already taken into account in the
NLO corrections to the ``one fragmentation" contribution and leading  order
``two fragmentation" components respectively.}, although they could also be
large. Secondly one should also even worry about the next correction to the
box, because the latter may be quite sizeable. Such a possibility is suggested
by the situation occurring to the first correction to the effective vertex $g g
\rightarrow h$, computed in \cite{dgsz}, and shown to reach generically about
50 \% of the  one-loop result. Moreover, this box contribution is the leading
order of a new mechanism, whose spurious (factorization and renormalization)
scale dependences are monotonic, and only higher order corrections would partly
cure this problem and provide a quantitative estimate. This tremendous effort
has not been carried out yet, although progresses towards this goal have been
achieved  recently \cite{delduca,deflorian,smirnov-tausk}.

\end{fmffile}

\subsection{Presentation of the method}

In  \cite{abfs}, a dedicated calculation was required for each observable.
Since then more versatile approaches have been developed, which combine
analytical and Monte-Carlo integration techniques \cite{owens}, \cite{cfg}.
They thus allow the computation of several observables within the same
calculation, at NLO accuracy, together with the incorporation of
selection/isolation cuts at the partonic level in order to match the various
cuts used by the experimental collaborations as faithfully as possible. The
studies of \cite{owens} and of \cite{yuan} rely on such an approach. Let us
briefly describe the one which we use here.

\subsubsection{Phase space slicing and subtraction of long distance
singularities}\label{phase-space-slicing}

Within the combined analytical and Monte-Carlo approach, two generic well known
methods can be used to deal with infrared and collinear singularities which are
met in the calculation of inclusive cross sections: the phase space slicing
method \cite{slicing} and the subtraction method \cite{subtraction}. The
approach followed in the present work uses a modified version of the one
presented in \cite{cfg}, which combines these two techniques.\\ 

For a generic reaction $1 + 2 \rightarrow 3 + 4 + 5$ two particles of the
final state, say 3 and 4, have a high $p_{T}$ and are well separated in phase
space, while the last one, say 5, can be soft, or collinear to either of the
four others. The phase space is sliced using two arbitrary, unphysical
parameters $p_{Tm}$ and $R$ in the following way:
\begin{itemize}
\item[-] Part I \\
The norm $p_{T 5}$ of transverse momentum of the particle 5 is required to be
less than some arbitrary value $p_{Tm}$ taken to be small compared to the other
transverse momenta. This cylinder supplies the infrared, and initial state
collinear singularities. It also yields a small fraction of the final state
collinear singularities. 
\item[-] Part II a \\
The transverse momentum vector of the particle 5 is required to have a norm
larger than $p_{Tm}$, and to belong to a cone $C_3$ about the direction of 
particle $3$, defined by  
$(y_5-y_3)^2+(\phi_5-\phi_3)^2~\leq~R_{th}^2$, with $R_{th}$ some small 
arbitrary number. $C_3$ contains the final state collinear singularities 
appearing when $5$ is collinear to $3$.
\item[-] Part II b \\
The transverse momentum vector of the particle 5 is required to have a
norm larger than $p_{Tm}$, and to belong to a cone $C_4$ 
about the direction of particle $4$, defined by 
$(y_5-y_4)^2+(\phi_5-\phi_4)^2~\leq~R_{th}^2$. $C_4$ contains the final 
state collinear singularities appearing when $5$ is collinear to $4$.
\item[-] Part II c \\
The transverse momentum vector of the particle 5 is required to have a
norm larger than $p_{Tm}$, and to belong to neither of the two cones $C_3$, 
$C_4$. This slice yields no divergence, and can thus be treated directly in
$4$ dimensions.
\end{itemize}

Collinear and soft singularities appear when integration over the kinematic
variables (transverse momentum, rapidity and azimuthal angles) of the particle
5 is performed on parts I, II a and II b. They are first regularized by
dimensional continuation from $4$ to $d = 4 - 2 \epsilon$, $\epsilon < 0$.  The
$d$-dimensional integration over the particle 5 on these phase space slices
yields these singularities as $1/\epsilon$ poles together with non singular
terms as  $\epsilon \rightarrow 0$. After combination with the corresponding
virtual contributions, the infrared singularities cancel, and the remaining
collinear singularities which do not cancel are factorized and absorbed in
parton distribution or fragmentation functions. The resulting quantities
correspond to pseudo cross sections where the hard partons  are unresolved from
the soft or collinear parton 5, which has been ``integrated out" inclusively on
the parts I, II a, II b. The word ``pseudo" means that they are not genuine
cross sections, as they are not positive in general. They are split into two
kinds. We call pseudo cross section for some $2 \rightarrow 2$ process the sum
of the lowest order term plus the fraction of the corresponding virtual
corrections where the infrared and collinear singularities have been
subtracted, and which have the kinematics of a genuine $2 \rightarrow 2$
process. The contributions where the uncanceled collinear singularities are
absorbed into parton distribution (on part I) or fragmentation (on parts II a
and II b) functions involve an extra convolution over a variable of collinear
splitting, as compared to the kinematics of a genuine $2 \rightarrow 2$
process: we call them pseudo cross sections for quasi $2 \rightarrow 2$
processes. The detailed content of these terms is given in the Appendix
\ref{ap}. For an extended presentation of the  details and corresponding
explicit formulas, we refer to \cite{cfg}.\\

As a matter of principle, observables do not depend on the unphysical
parameters $p_{Tm}$ and $R_{th}$. Yet, the pseudo cross sections on parts I, 
II a, II b and II c separately do. Let us briefly discuss the cancellation of
the $p_{Tm}$ and $R_{th}$ dependences in observables computed according to this
method. In the cylindrical part I, the finite terms  produced are approximated
in order to collect all the terms depending logarithmically on $p_{Tm}$,
whereas terms proportional to powers of $p_{Tm}$ are neglected. This differs
from the subtraction method implemented in the cylinder in \cite{cfg}, which
kept the exact $p_{Tm}$ dependence. On the other hand, in the conical parts II
a and II b,  the same subtraction method as in \cite{cfg} is used, so that the
exact $R_{th}$ dependence is kept. This ensures the exact cancellation of the 
dependence on the unphysical parameter $R_{th}$  between part II c and parts II
a, II b whereas only an approximated cancellation of the unphysical parameter
$p_{Tm}$ dependence between parts II c, II a and  II b and part I occurs. The
parameter $p_{Tm}$ must be chosen small enough with respect to $p_{T3}$ and
$p_{T4}$ in order that the neglected terms can be safely dropped out. In
practice, it has been verified that $p_{Tm}$ values of the order of half a
percent of the minimum of $p_{T3}$ and $p_{T4}$ fulfill these requirements. A
more detailed discussion on this issue  is provided in Appendix
\ref{appendix-b}.\\

The pseudo cross sections on parts I, II a, II b, as well as the transition
matrix elements on the part II c, are then used to sample unweighted kinematic
configurations, in the framework of a partonic event generator, described in \ref{eventgen} below. 

\subsubsection{Partonic event generator}\label{eventgen}

For practical purposes, a partonic event generator has been built for diphoton
production including all the mechanisms: the ``direct", ``one-" and ``two
fragmentation". Each mechanism is treated separately. Firstly, the contribution
of a given mechanism to the integrated cross section is calculated with the
integration package BASES \cite{bases}. At this stage, some kinematic cuts
(e.g. on the rapidity of the two photons, on their transverse momenta, etc.)
may be already taken into account.  Then, for the $2 \rightarrow 2$
contributions and the quasi $2 \rightarrow 2$ contributions on the parts I, II
a and II b, and the inelastic contributions on the part II c of the phase
space, partonic events are generated with SPRING \cite{bases} with a weight
$\pm 1$ depending on the sign of the integrand at this point of the phase
space\footnote{This trick circumvents the fact that SPRING works only with
positive integrands, while the pseudo cross sections are not positive. The
generated events are thus unweighted up to a sign.}. All the events are
subsequently stored into a NTUPLE \cite{paw}. Finally these NTUPLES can be
histogramed at will, incorporating any further cuts, such as those imposed by
some isolation criterion as discussed in the next subsection. It is suitable
to use values for $R_{th}$ and $p_{Tm}$ which are fairly small and
disconnected from any physical parameter. The phase space generation is then as
exclusive as possible. Moreover it allows to investigate the dependence of
various observables with respect to the physical isolation parameters, as well
as to investigate different types of isolation criteria, using an event sample
conveniently generated {\it once for all}.
In practice however one
cannot use too small values in order to keep statistical fluctuations under
control, unless the computer time and the sizes of the NTUPLES become
intractably large.\\

Let us state more clearly what we mean by {\it partonic event generator}. Since
the events associated to the $2 \rightarrow 2$ and quasi $2 \rightarrow 2$ 
contributions have a negative weight, this code, properly speaking, is not a 
genuine event generator on an event-by-event basis. By events, we mean final
state partonic configurations. For a given event, the informations stored into
the NTUPLE are the 4-momenta of the outgoing particles; their flavors: parton
(i.e. quark or gluon) or photon; in the fragmentation cases, the longitudinal
fragmentation variable(s) associated with the photon(s) from fragmentation;
and, for practical purpose, a label which identifies the type of pseudo cross
section ($2 \rightarrow 2$, quasi $2 \rightarrow 2$, inelastic) which produced
the event stored. Notice also that in the fragmentation cases, all but the
longitudinal information on the kinematics of the residue of the collinear
fragmentation is lost. Hence this type of program does not provide a realistic,
exclusive portrait of final states as given by genuine, full event generators
like PYTHIA \cite{pythia} or HERWIG \cite{herwig}. On the other hand, the
latter are only of some improved leading logarithmic accuracy. Thus, our code is
more precisely a general purpose computer program of Monte-Carlo type, whose
virtue is the computation of various inclusive enough observables within the 
same calculation, at NLO accuracy.

\subsection{The implementation of isolation cuts}

Collider experiments at $Sp\bar{p}S$, the Tevatron, and the forthcoming LHC do
{\it not} measure {\it inclusive} photons. Indeed, the inclusive production 
rates of high $p_{T}$ $\pi^{0}$, $\eta$, $\omega$, or of pairs  $\pi^{0}
\pi^{0}$ or $\gamma \pi^{0}$, etc, with large invariant mass, are orders of
magnitudes larger than for direct photons. In order to reject the huge
background of secondary photons produced in the decays of these mesons, the experimental event selection of direct photons (single photons, as
well as diphotons) requires the use of isolation cuts. Such a requirement will
be absolutely crucial at LHC for the search of Higgs bosons in the two photon
channel and the mass range 90-140 GeV, since the expected background from
$\pi^{0}$, etc. is about eight orders of magnitudes larger than the signal
before any isolation cut is applied. \\

A widely used criterion to isolate photons is schematically the
following\footnote{An alternative to the criterion (\ref{criterion}) has been 
recently proposed in \cite{frixione}, in which the veto on accompanying
hadronic transverse energy is the more severe, the closer the corresponding
hadron to the photon direction. It has been designed to make the
``fragmentation" contribution vanish in an infrared safe way.}. A photon is
said to be isolated if, inside a cone centered around the photon direction in the rapidity and azimuthal angle plane, the amount of hadronic transverse energy
$E_T^{had}$ deposited is smaller than some value $E_{T\, max}$ fixed by the
experiment:
\begin{equation}\label{criterion}
\left.
\begin{array}{rcc} 
\left(  y - y_{\gamma} \right)^{2} +  \left(  \phi - \phi_{\gamma} \right)^{2}  
& \leq  & R^{2} \\
E_T^{had} & \leq & E_{T\, max}
\end{array}
\right\} 
\end{equation} 

The topic of the isolation of photons
based on the above cone criterion (\ref{criterion}) has been rather extensively 
discussed in the theoretical literature, especially in the case of production
of single isolated photons in hadronic\footnote{The related topic of isolated
prompt photons produced in $e^{+}e^{-}$ annihilation into hadrons has also been
abundantly discussed. A variant of the definition (\ref{criterion}) suitable
for $e^{+}e^{-}$ has been studied in \cite{Kunszt-Troscanyi}, and recently
revisited in \cite{berger-guo-qiu,cfp}. An alternative criterion has been
proposed in \cite{glover-morgan}, and applied to the measurement of isolated
photons in LEP experiments.} collisions
\cite{boo1,abf,berger-qiu,gordon-vogelsang,cfgp}.
Beside the rejection of the background of secondary photons, the isolation
requirement also reduces the photons from fragmentation. The account of
isolation effects on the ``fragmentation" contribution was accurate to LO
accuracy in \cite{boo1,abf}. A treatment to NLO accuracy has been subsequently
given in \cite{gordon-vogelsang}, following the subtraction framework
presented in \cite{berger-qiu}. Isolation implies however that one is not
dealing with inclusive quantities anymore. This raised questions concerning
the validity of the factorization property in this case, and whether the 
fragmentation functions may depend on the isolation parameters, as  assumed
in \cite{berger-qiu}. This raised also issues regarding soft gluon divergences
in isolated photons cross sections, as in \cite{berger-guo-qiu}. These
questions have been clarified in \cite{cfgp,cfp}. The factorization property
of collinear singularities still holds for cross sections based on 
the criterion\footnote{The fact that {\it transverse} energies are involved in
(\ref{criterion}) in hadronic collisions is crucial in this
respect. Factorization would be broken if energies were used instead.}, and the
fragmentation functions involved there are the same as in the inclusive case,
whereas the effects of isolation are consistently taken into account in the
short distance part. Yet cross sections defined with this criterion may have
infrared divergences - or, at least, instabilities, depending on the
inclusiveness of the observable considered - located at some isolated critical
points {\it inside} the physical spectrum of some observables calculated at
fixed order, namely NLO, accuracy. This means that the vicinities of these
critical points are sensitive to multiple soft gluon effects, which have to be
properly taken into account in order to provide correct predictions.\\

In the present calculation, as in \cite{gordon-vogelsang,cfgp} for the case of
single photon production, the transverse energy deposited in the cone may come
from the residue of the fragmentation, from the parton $5$ (which never
fragments into photons) or from both. During the projection of the NTUPLES onto
any desired observable, the isolation criterion (\ref{criterion}) about the
two photons is applied to each stored partonic configuration. The effects of
isolation are commented in \ref{effects_isolation}. In addition, at
the NLO accuracy at which our calculation is performed, potentially large
logarithmic contributions of infrared origin may be induced by the extra
isolation constraint on the phase space. The issue of infrared
sensitivity induced by isolation will be discussed further in 
\ref{inside}. Let us mention that no summation of such logarithms is
performed in our treatment.

\section{Phenomenology}\label{phenom}

In this section, we adopt a LHC oriented presentation. We start with a brief
comparison of our NLO calculations with WA70 and D0 data for illustrative
purposes. We then show some predictions for LHC in the invariant mass range $80$
GeV$\leq m_{\gamma \gamma} \leq 140$ GeV corresponding to the Higgs boson
search through $h \to \gamma \gamma$. We discuss the ambiguities plaguing
these predictions due to the arbitrariness in the choices of the
renormalization scale $\mu$, of the  initial state factorization scale $M$
(which enters in the parton distribution  functions), and of the fragmentation
scale $M_{f}$.

\subsection{Comparison with experimental data}

\subsubsection{Fixed target data}

A comparison between the diphoton differential cross section versus each
photon's transverse momentum measured by the WA70 collaboration
\cite{wa70cross} and our NLO postdiction is shown on Fig.~\ref{flc}, together 
with the respective magnitude of the various contributions. The NLO
calculation has been made with the ABFOW parton distribution
functions~\cite{abfow} for the proton and the corresponding ones for the
pion~\cite{abfkw}\footnote{The choice of the parton distributions is mandated
by the  fact that  the initial state of the reaction is $\pi^{\pm}$ proton.
Therefore a  consistent set of parton densities inside the proton and the pion
must be  taken. Indeed, to extract the parton distribution functions in the
pion,  reactions such as $\pi^{\pm} \; p \rightarrow \gamma^* \, X$ (Drell-Yan)
and  $\pi^{\pm} \; p \rightarrow \gamma \, X$ (direct photon) are used.
Consequently  some correlations between the proton and the pion partonic 
densities exist, and it is preferable to use consistent sets in the
calculations. Only three groups provided such a work: ABFW \cite{abfkw}, MRS 
\cite{smrs} and GRV \cite{grv}. All these works are rather old and the 
partonic densities are rather similar in the WA70 x range.}, for the scale
choice\footnote{One shall not attach importance to the somewhat unusual value 
$\lambda=0.275$ of the scale choice. Relatively low scales such as this one, or
$\lambda=0.25$ equally well, turn out to match the data better than higher
scale choices. Yet this particular value was not chosen as the one which
matches the data the best, but for a minor though cumbersome computational
reason. The WA70 collaboration requires the  transverse momenta of the photons
to be larger than $3$ GeV and $2.75$ GeV respectively. However for
computational convenience we first implemented a symmetric cut on the $p_{T}$
of each photon: $p_{T}\geq 2.75$ GeV at the level of the Monte Carlo generation
of photon pairs.  In the ABFOW parametrizations, the factorization scale
$M^{2}$ has to be larger than $2$  GeV$^{2}$. Given the above symmetric cut on
both photons  in the Monte Carlo generation, taking $\lambda = 1/4$ does not
ensure that  $M^{2}$ is always above $2$ GeV$^{2}$, while the choice $\lambda
\geq .275$  does.}  $M = M_{f} = \mu = \lambda (p_{T}(\gamma_{1}) +
p_{T}(\gamma_{2}))$, with   $\lambda=0.275$. The ``one fragmentation"
contribution is one order of magnitude below the ``two direct" contribution.
The ``two fragmentation" contribution is even smaller and negligible here. The
smallness of these contributions is the reason why previous works
\cite{abfs,owens} described this observable reasonably well too, despite the
absence of higher order corrections to the fragmentation contributions there.\\

Various correlations between the two photons: the distribution of the $p_{T}$ 
imbalance variable $z = -{\mbox{\bf p}}_{T}(\gamma_{1}).{\mbox{\bf p}}_{T}
(\gamma_{2}) /p^{2}_{T}(\gamma_{1})$   the distribution of the azimuthal angle
between the two photons ($\phi_{\gamma \gamma}$), the distribution of $p_{
out}$\footnote{The beam axis together with the direction of one of the two
photons define a plane. The component of the transverse momentum of the other
photon along the direction perpendicular to this plane is the $p_{out}$ of this
photon.}, and the distribution of  transverse momentum of diphotons
($q_{T}$), have been measured also by the  WA70 collaboration
\cite{wa70correl}. These distributions are infrared sensitive near the elastic
boundary of the spectrum (e.g. $q_{T} \rightarrow 0$ or $\phi_{\gamma \gamma}
\rightarrow \pi$) or near a critical point (e.g. $z = 1$) and, moreover, are
quite sensitive to non perturbative  effects appearing in the resummed part of
calculations summing soft gluon effects. This sensitivity extends over a wide
part of the spectrum covered by the measurements. Consequently we do not
present any comparison of these data points with the approximation of fixed
order accuracy of this work; nor will we discuss the scale ambiguities at fixed
target energies.

\subsubsection{Tevatron collider data}

A preliminary study of diphotons events in the central region
($|y(\gamma_{1,2})| < 1.0$) has been recently performed by the D0 collaboration
\cite{d0}.\\ 

\noindent
The experimental cuts in the D0 data used for the comparisons are not corrected
for electromagnetic calorimeter absolute energy scale. The electromagnetic
energy scale correction is given by \cite{d0}:
\[
E(\mbox{measured}) = \alpha \; E(\mbox{true}) + \delta
\]
where 
\begin{eqnarray}
\alpha & = & 0.9514 \pm 0.0018 {}^{+0.0061}_{-0.0017} \nonumber \\
\delta & = & -0.158 \pm 0.015  {}^{+0.03}_{-0.21} \mbox{GeV} \nonumber
\end{eqnarray}
Thus, the experimental cuts at measured values of 14 (respectively 13) GeV 
correspond to cuts at roughly 14.90 (resp. 13.85) GeV in the theoretical
calculation. Smearing effects accounting for electromagnetic calorimeter
resolution have not been implemented, but given the experimental fractional
energy resolution of the electromagnetic calorimeter \cite{Wmass-d0}, they are
expected to be of the level of a few  percent only.\\

The actual isolation cuts used experimentally (such as vetoes on charged tracks
in some conical vicinity about each photon, etc.) are quite more complicated
than the schematic criterion (\ref{criterion}), and cannot be faithfully
implemented at the partonic level. We instead simulated them in our NLO
calculation by requiring that the accompanying transverse partonic energy be
less than $E_{T \, max} = 2$ GeV in a cone $R = 0.4$ about each photon. Varying
$E_{T \, max}$ from $1$ to $3$ GeV in the calculated cross-section, as a rough
estimate of the effects of smearing due to hadronic calorimeter resolution and unfolding of underlying events contribution turns out to have 
a less than 4\% effect.\\

The MRST2 set of parton distributions functions\footnote{The MRST sets 1,2,3 
are associated with the value $\Lambda_{\overline{MS}}=300$ MeV for $n_{f} = 4$ 
flavors. This corresponds to $\alpha_s(m_Z) = 0.1175$ in the $\overline{MS}$ scheme. For more details, see \cite{mrs99}} \cite{mrs99} is
used\footnote{The MRST1 set is presented by the authors of \cite{mrs99} as the
default set. However, in order to take into account mutually inconsistent data
sets on single direct photon production at fixed targets, a $k_{T}$ smearing
procedure is involved in the determination of this set. This procedure is
strongly model dependent and questionable as long as no unambiguous way is
found to lodge it in the QCD improved parton model. The set MRST2 does not
involve this procedure, so we prefer to base any prediction and comparison on
this set.}, with the scales arbitrarily chosen to be $M = M_f = \mu =m_{\gamma
\gamma}/2$.  The prediction for the above scale choice is shown for the
diphoton differential cross sections vs. the transverse momentum of each
photon  (Fig.~\ref{flc1}), the diphoton mass (Fig.~\ref{flc2}), and for the
transverse momentum of photon pairs  (Fig.~\ref{flc3}) and the azimuthal angle
between the photons  (Fig.~\ref{flc4}). With the scale choice used, the ``one fragmentation" contribution is roughly one tenth of the ``direct" one whereas the ``two fragmentation" yields a tiny contribution. To illustrate this, the
different contributions: ``direct", ``one-" and ``two  fragmentation" are shown
separately on Fig.~\ref{flc4}.  The distributions of the transverse momentum of photon
pairs and of the azimuthal angle between the photons are well known to be
controlled by multiple soft gluon emission near the elastic boundary of the
spectrum, $q_{T} \rightarrow 0$ and $\phi_{\gamma \gamma} \rightarrow \pi$
respectively. Consequently, the accuracy of any fixed-order calculation,
including the present one, is not suited to study such observables in these
respective ranges. More on this issue will be commented in the next section. On
the other hand a NLO calculation is expected to be predictive for the tails of
these distributions away from the infrared sensitive region.\\

The data are reasonably described, taking into account a correlated
systematic error for events in which the $p_T$ of both photons is above 20
GeV. This correlated systematic error due to the background evaluation affects
obviously the three highest $p_T$ points of the  transverse energy spectrum, as
well as the three highest points of the diphoton mass spectrum. \\ 

We do not present any analysis of the various scale dependences for Tevatron.
Such a discussion is proposed for LHC in the next section. Yet let us mention
that, at Tevatron, the energy scale is lower and the relevant values of $x$
are somewhat higher than at LHC. Consequently, the renormalization scale
dependence is slightly sharper, on the other hand the factorization scale
dependence is somewhat flatter than at LHC. Nevertheless the situation at
Tevatron is expected to be qualitatively similar to the one at LHC.

\subsection{Predictions for LHC}

We now discuss some results computed with the kinematic cuts from the CMS 
and ATLAS proposals \cite{cms}, namely $p_T(\gamma_1)> 40$  GeV, 
$p_T(\gamma_2)> 25$ GeV, $|y(\gamma_{1,2})| < 2.5$, with $80$ GeV $\leq 
m_{\gamma \gamma} \leq 140$ GeV, and using the MRST2 set of parton distribution
functions \cite{mrs99} and the fragmentation functions of  \cite{phofrag}.

\subsubsection{Scale ambiguities}\label{scaledep}

We first consider the invariant mass distribution of diphotons, in absence of 
isolation cuts, cf. Fig.~\ref{flc5} in order to illustrate the
strong dependence of the splitting into the three contributions, ``direct",
``one-" and ``two fragmentation",  on the scale chosen, as we warned in~\ref{theorcont}. 
In both choices of scales displayed the ``one fragmentation" contribution 
dominates, but the hierarchy between ``direct" and "two fragmentation" 
contributions is reversed from one choice to the other. With the choice of scales $M = M_f = \mu = m_{\gamma \gamma}/2$, the 
``one fragmentation" is more than twice larger than the ``direct" one, and
the ``two fragmentation" is the smallest. On the other hand, with the other 
choice $M = M_f = \mu = 2 m_{\gamma \gamma}$, the ``one fragmentation" 
contribution is three to five times larger than the ``two fragmentation" 
component, and more than one order of magnitude above the ``direct" 
one. On the other hand the total contribution seems rather stable.\\

Yet the arbitrariness in the choices of the various scales still induces theoretical
uncertainties in NLO calculations. In the following we actually do not perform 
a complete investigation of all three scale ambiguities independently with 
search for an optimal region of minimal sensitivity. At the present stage, we 
limit the study to an estimation of the pattern and magnitude of their effect 
on our results. We show how the scale ambiguities affect our prediction for the 
invariant mass distribution. We consider both the case without isolation (Fig.
\ref{fig.scale.dep.incl}) and the isolated case with $E_{T \, max} = 5$ GeV 
inside $R \leq 0.4$ (Fig. \ref{fig.scale.dep.isol}). For the present purpose, 
the virtue of the actual values of the isolation parameters used here is to 
strongly suppress the fragmentation contributions hence the associated 
$M_{f}$ dependence. We compare four different choices of scales: two choices 
along the first diagonal $\mu = M = M_{f} =  m_{\gamma \gamma}/2$ and  
$\mu = M = M_{f} = 2 \, m_{\gamma \gamma}$; and two anti diagonal choices, 
$\mu =  m_{\gamma \gamma}/2  ; M = M_{f} = 2 \, m_{\gamma \gamma}$ and 
$\mu = 2 \, m_{\gamma \gamma} ; M = M_{f} = m_{\gamma \gamma}/2$. We do not 
perform a separate study of fragmentation scale dependence. Yet the latter can 
be indirectly estimated by comparing the  results of the isolated case, where
the fragmentation components, thereby the corresponding fragmentation scale
dependence, are strongly suppressed, with the  situation in the non isolated
case, where especially the ``one fragmentation" contribution is quite large,
and the ``two fragmentation" not negligible, so that the issue of fragmentation
scale dependence matters. \\

When scales are varied between $m_{\gamma \gamma}/2$ and $2 \, m_{\gamma
\gamma}$  along the first diagonal $\mu = M = M_{f}$, the NLO results for the
invariant  mass distribution appear surprisingly stable, since they change by
about 5\% only. Alternatively, anti-diagonal variations of $\mu$ and $M =
M_{f}$ in the  same interval about the central value $m_{\gamma \gamma}$ lead
to a variation  still rather large (up to 20 \% cf. Fig.
\ref{fig.scale.dep.incl} and Fig. \ref{fig.scale.dep.isol}). This is because variations
with respect to $\mu$ and $M$ are separately monotonous but act in opposite ways.
When  $\mu$ is increased, $\alpha_{s}(\mu^{2})$ hence the NLO corrections 
decrease\footnote{ In processes for which the lowest order is proportional to 
some power $\alpha_{s}^{n}, n \geq 1$, an explicit $\mu$ dependence appears in
the next-to-leading order coefficient function, which partially compensates the
(large) $\mu$ dependence in $\alpha_{s}(\mu^{2})$ weighting the lowest  order.
Unlike this, in the ``two direct" component which dominates the  cross section
when a drastic isolation is required, the lowest order  involves no
$\alpha_{s}$. This leads to a rather small $\mu$ dependence, since the latter
starts only at NLO. On the other hand, the $\mu$  dependence occurs only
through the monotonous decrease of the  $\alpha_{s}(\mu^{2})$ weighting the
first higher order correction: there is no  partial cancellation of $\mu$
dependence. Such cancellation would start only at  ${\cal O}(\alpha_{s}^{2})$,
i.e. at NNLO.  The mechanism is more complicated in presence of fragmentation
components, and the situation becomes mixed up between all components when the
severity of isolation is reduced.}. On the other hand the relevant values of
momentum fraction of incoming partons are small, $\sim {\cal O}(10^{-3}$ to
$10^{-2})$ so that the gluon and sea quark distribution functions increase when
$M$ is increased. In the isolated case, this leads to a monotonous increase of
the ``direct" component, over a large band of the invariant mass range
considered, as $M$ is increased, cf. Fig. \ref{Mmonotonous}, which is induced
in particular by the monotonous increase of the box contribution.  Scale
changes with respect to $\mu$ and $M$ turn out to nearly cancel against each
other along the first diagonal but add up in the other case. Actually, the
stability along the first diagonal is accidental.\\

In conclusion, the $\mu$, $M$ dependences are thus not completely under control
yet at NLO in the kinematic range considered. On the opposite, the account
for  the NLO corrections to the fragmentation components provides some 
stability with respect to $M_{f}$ variations about orthodox choices of the
fragmentation scale. \\

The issue of $\mu$ dependence of less inclusive observables, such as the tails 
of the $q_{T}$ or $\phi_{\gamma \gamma}$ distributions are the same for the
invariant mass distribution. This is because the tails of these distributions 
is purely given by the NLO corrections and dominated by the  ${\cal
O}(\alpha_{s})$ corrections of the ``two direct" component. On the other hand,
the $M$ dependence is a bit larger, so is the combined uncertainty on the
theoretical results for these distributions, cf. Fig. \ref{qt_lhc_5GeV} and
Fig. \ref{phi_lhc_5GeV}.\\

\subsubsection{Effect of isolation}\label{effects_isolation}

We now consider the effect of isolation on the various contributions. As
expected, isolation reduces the diphoton production rate, with respect to the
inclusive case, cf. Fig. \ref{incliso}. More precisely, severe isolation requirements like 
$E_{T \, max} = 5$~GeV inside a cone $R = 0.4$ suppress the ``one
fragmentation" component, which dominates the inclusive rate, by a factor 20 to
50, and kill the ``two fragmentation"  contribution completely. \\

However this net result hides a rather intricate mechanism, cf. Fig. 
\ref{flc55} vs. Fig. \ref{flc5}, by which the ``two direct" contribution turns
out to be increased! Surprising as it may seem at first sight, this effect has
the following origin. Higher order corrections  to the ``two direct" component
involve in particular the two  subprocesses $\bar{q}q \rightarrow \gamma \gamma
g$ and  $g q \rightarrow \gamma \gamma q$ (where $q$ is a quark or an
antiquark).  The first one yields a positive contribution. On the other hand,
the collinear  safe part of the second one  yields a contribution which is
negative, and  larger in absolute value than the previous one in the inclusive
case, as was already seen in \cite{abfs}.  Isolation turns out to suppress more
the higher order corrections from the second mechanism than from the first one,
so that the NLO isolated ``two direct" contribution is larger than the
inclusive one. Yet, the ``fragmentation" contributions are suppressed more than
the ``two direct" one is increased, so that the sum of all contributions is
indeed decreased, with respect to the inclusive case. Once again, one has to
remember that the splitting into the three mechanisms depend, not only on the
factorization scale, but more generally on the factorization scheme.  This
arbitrariness generates such counterintuitive offsprings; in a final state 
factorization scheme different than the ${\overline{MS}}$ scheme, the various 
components, especially the ``two direct" one, may be separately affected by
isolation cuts in a different way. This once more illustrates the danger of
playing with these unphysical quantities separately.\\

A more detailed analysis of the dependence of NLO estimations of various
observables on the isolation cut parameters, especially on $E_{T \, max}$ will 
be given in a forthcoming publication. We will also come back to this issue,
regarding infrared sensitivity, in~\ref{stringent}.\\

\section{Infrared sensitive observables of photon pairs and soft
gluon divergences.}\label{irsensible}

Being based on a fixed, finite order calculation, our computer code is not
suited for the study of observables controlled by multiple soft gluon
emission, and has to be improved in this direction. Among these infrared 
sensitive observables, one may distinguish the following examples, most of 
which would require an improved account of soft gluon effects.

\subsection{Infrared sensitivity near the elastic boundary}

\subsubsection{The transverse momentum distribution 
$d \sigma/ d q_{T}$ of photon pairs near $q_{T} = 0$} 

Both in the inclusive and isolated cases, this distribution is an infrared 
sensitive observable, controlled by the multiple emission of soft and 
collinear gluons. This well known phenomenon has been extensively studied for
the corresponding observable in the Drell-Yan  process \cite{reviewir}. A loss
of balance  between the contribution of real emission, strongly suppressed near
this  exclusive phase space boundary, and the corresponding virtual
contribution, results in large Sudakov-type logarithms of $m^{2}/q_{T}^{2}$
($m$  being the invariant mass and and $q_{T}$ the transverse momentum of the
photon  pair - the heavy vector boson in the Drell-Yan case) at every order in
perturbation. In order to make sensible predictions in this regime, these
Sudakov-type logarithms have to be resummed to all orders. \\

The treatment of the ``two direct" and box contributions is similar to the 
well-known Drell-Yan process, and has been carried out recently by 
\cite{balacz} at next-to-leading logarithmic accuracy in the framework tailored
by Collins, Soper and Sterman \cite{cs}. On the other hand, the  fragmentation
contributions do not diverge order by order when  $q_{T} \rightarrow 0$.
Indeed, in the ``one fragmentation" case,
\begin{eqnarray} 
\mbox{parton}_{1} + \mbox{parton}_{2} & \rightarrow & 
     \gamma_{1} + \mbox{parton}_{3}  \label{hard1}  \\
\mbox{parton}_{3} & \rightarrow &  \gamma_{2} + X \label{hard2} 
\end{eqnarray}
the NLO contribution to the hard subprocess (\ref{hard1}) 
yields a double logarithm of the form
\begin{equation}
\sim \alpha_{s} \ln^{2} \| {\mbox {\bf p}}_{T}(\gamma_{1}) + 
{\mbox {\bf p}}_{T}(\mbox{parton}_{3}) \| 
\end{equation}
when $\|{\mbox {\bf p}}_{T}(\gamma_{1}) + 
{\mbox {\bf p}}_{T}(\mbox{parton}_{3})\| \rightarrow 0$. 
However the extra convolution associated with the fragmentation
(\ref{hard2}) involves an integration over the fragmentation variable
$p_{T}(\gamma_{2})/p_{T}(\mbox{parton}_{3})$ which 
smears out this integrable singularity. The ``two fragmentation" 
contribution involves two such convolutions, hence one more smearing.\\

\subsubsection{The distribution of photon-photon azimuthal angle
$d \sigma/d \phi_{\gamma \gamma}$ near $\phi_{\gamma \gamma} = \pi$} 

This distribution is another interesting infrared sensitive observable,
measured by several experiments both at fixed target and collider energies
\cite{wa70correl,cdf1,d0}, though less discussed in the literature from the
theoretical side. The regime $\phi_{\gamma \gamma} \rightarrow \pi$ includes
back-to-back photons, a set of configurations which lie at the elastic boundary
of the phase space. This case differs from the previous one for two reasons.
Firstly, not only the ``two direct" contribution diverges order by order when
$\phi_{\gamma \gamma} \rightarrow \pi$, but also both ``one-" and  ``two
fragmentation" contributions diverge as well, as can be inferred from Fig.
\ref{flc4}. Indeed, consider the example of the ``one fragmentation case", cf.
equations \ref{hard1} and \ref{hard2}. Selecting  $\phi_{\gamma \gamma}
\rightarrow \pi$ emphasizes the configurations with  $\phi(\mbox{parton}_{3}) -
\phi(\gamma_{1}) \rightarrow \pi$,  so that all the emitted partons besides
parton 3 have to be collinear  to either of the incoming or outgoing particles,
and/or soft, which  yields double logarithms
\begin{equation}
\sim \alpha_{s} \ln^{2} \left[ \pi - 
\left( \phi(\mbox{parton}_{3}) - \phi(\gamma_{1}) \right) \right]
\end{equation}
associated with each of the hard partons $1,2,3$ - plus single
logarithms as well. For the observable 
$d \sigma / d \phi_{\gamma \gamma}$ near
$\phi_{\gamma \gamma} = \pi$, the integral involved in the convolution of the
hard subprocess with the fragmentation functions does not smear these
logarithmic divergences, since the fragmentation variable  
$p_{T}(\gamma_{2})/p_{T}(\mbox{parton}_{3})$ is 
decoupled from the azimuthal variable 
$\phi(\mbox{parton}_{3})$ which is equal to $\phi(\gamma_{2})$, $\gamma_{2}$ 
and parton $3$ being collinear. A similar observation holds for the ``two
fragmentation" component. Moreover, in both fragmentation cases, 
soft gluons may couple to both initial and final state hard emitters. The 
resulting color structure of
the emitters is more involved than in the ``two direct" case, and especially 
more complicated in the ``two fragmentation" case as shown in some recent works
\cite{sterman-kidonakis}. This would make any resummation 
quite intricate beyond leading logarithms.\\

Let us notice that both fragmentation components make 
$d \sigma / d \phi_{\gamma \gamma}$ diverge also when
$\phi_{\gamma \gamma} \rightarrow 0$. The increase of the fragmentation 
contributions in the lower $\phi_{\gamma \gamma}$ range 
is the trace of this divergence, cf. Fig. \ref{flc4}.

\subsection{An infrared divergence {\it inside} the physical
region.}\label{inside}

In the case of photons isolated with the standard fixed cone size criterion of 
eqn. (\ref{criterion}), a new problem appears in the $q_{T}$ distribution. This
problem  does not concern the region $q_{T} \rightarrow 0$; still it has to do
with  infrared and collinear divergences. This can be seen on Fig.
\ref{flc13},  which shows the observable $d \sigma/d q_{T}$ vs. $q_{T}$ for
isolated photon  pairs, computed at NLO accuracy. The computed $q_{T}$
distribution turns out to  diverge   when $q_{T} \rightarrow E_{T \, max}$ from
below. Notice that the critical point $E_{T \, max}$ is located {\it inside}
the physical region. The phenomenon is similar to the one discovered in 
\cite{berger-qiu} in the production of isolated photons in $e^{+}e^{-}$
annihilation, and whose physical explanation has been given in 
\cite{cfp} following the general  framework of 
\cite{catani-webber}. It is a straightforward exercise to see that the lowest
order ``one fragmentation" contribution has a stepwise  behaviour, as noticed
in \cite{yuan}. Indeed, at this order,  the two photons are
back-to-back. $E_{T \; had}$ being the transverse hadronic  energy deposited in
the cone about the photon from fragmentation, the  conservation of transverse
momentum implies at this order that  $E_{T \, had} = q_{T}$. The corresponding
contribution to the differential cross section  $d \sigma/d q_{T}$ thus takes
the schematic form:
\begin{equation}
\left( \frac{d \sigma}{d q_{T}} \right) ^{(1 \; fragm, \;LO)} = 
 f \left( q_{T} \right) \; \Theta \left( E_{T \, max} - q_{T} \right)
\end{equation}
According to the general analysis of  \cite{catani-webber}, the  NLO
correction to $d \sigma/d q_{T}$  has a double logarithmic divergence at the
critical point $q_{T} = E_{T \, max}$ \footnote{In practice, the $q_{T}$
spectrum is  sampled into bins of finite size, and the distribution represented
on Fig. \ref{flc13} is averaged on each bin. Since the logarithmic
singularity is integrable, no divergence is actually produced. However when the
bin size is shrunk, the double logarithmic branch appears
again.}\label{nodiv}. The details of this infrared structure are very sensitive
to the kinematic constraints and the observable considered. In the case at
hand, at NLO, $d \sigma/d q_{T}$ gets a double logarithm {\it below} the
critical point, which is produced by the convolution  of the lowest order
stepwise term above, with the probability  distribution for emitting a
soft and collinear gluon:
\begin{eqnarray}
\left( \frac{d \sigma}{d q_{T}} \right) ^{(1 \; fragm, \;HO)} & \simeq &
- f \left( q_{T} \right) \; \Theta \left( E_{T \, max} - q_{T} \right) \nonumber \\
& & \mbox{} \times \frac{\alpha_{s}}{2 \pi} \; C \; 
\ln^{2} \left( 1 - \frac{q_{T}^{2}}{E_{T \, max}^{2}} \right) \;
 + \cdots
\end{eqnarray}
where $C$ is a color factor, $C_{F}$ or $N_{c}$ according to whether the
soft collinear gluon emitter is a quark (antiquark) or a gluon.
More generally, at each order in $\alpha_{s}$, up to two powers of such
logarithms will appear, making any fixed order calculation diverge at 
$q_{T} = E_{T \, max}$, so  that the spectrum computed by any fixed order
calculation is unreliable  in the vicinity of this critical value. An all order
resummation has to be carried out if possible in order to restore 
any predictability. A correlated step appears also in the ``two direct"
contribution at NLO, in the bin about $q_{T} = E_{T \, max}$. A detailed
study of these infrared  divergences will be presented in a future article.\\

No such divergence appears in the $q_{T}$ distribution of 
photon pairs presented in \cite{yuan}. The non 
appearance of the double logarithmic divergence there comes from the 
fact that the latter pops out only at NLO, while 
the authors of \cite{yuan} compute the ``one 
fragmentation" component at lowest order. 
Furthermore, the stepwise lowest order ``one fragmentation" contribution 
to the $q_{T}$ distribution is replaced in 
\cite{yuan} by the result of the Monte Carlo simulation 
of this component using PYTHIA \cite{pythia}. A quantitative comparison
is thus difficult to perform\footnote{Such a comparison involves two issues. \\
The first aspect concerns the infrared sensitivity below the critical
point. When the scale of $\alpha_{s}$ in the Sudakov
factor of the fragmenting quark is chosen to be the transverse momentum
of the emitted gluon with respect to the emitter, the parton shower not
only reproduces the fragmentation function of a parton into a photon to the collinear leading logarithmic approximation, but
it also provides an effective resummation of soft gluons effects to
infrared and collinear leading logarithmic accuracy. (This would not be true if,
instead, the scale of $\alpha_{s}$ in the Sudakov factor were the
virtuality of the emitter). This ensures that the distribution does not
diverge from below at the critical point, but rather tends to a finite
limit. \\
The second issue concerns the shape of the tail of this contribution above the
critical point. Indeed, energy-momentum conservation at each branching makes
the parton shower generate also contributions in the region  
$q_{T} > E_{T \, max}$, which is forbidden at lowest order. These
contributions  would be classified in a beyond  leading order calculation as
higher order corrections. Unlike in a fixed order calculation however, they 
provide only a partial account of such corrections, but to arbitrary high 
order. The accuracy of these terms is thus uneasy to characterize, and a
quantitative comparison between PYTHIA and any fixed order calculation is
difficult to perform.}.   \\

It can be noticed that the divergence at $q_{T} = E_{T \, max}$ is not visible
on Fig. \ref{flc3}. This is because in this case, the critical point $E_{T \,
max}$ in the $q_{T}$ spectrum where the theoretical calculation diverges is too
close to the other singular point $q_{T} = 0$, given the binning used.  The two
singularities contribute with   opposite signs in these bins and a numerical
compensation  occurs, resulting in no sizeable effect.  Yet the problem is only
camouflaged. A similar smearing appears also at  LHC energies for a stringent
isolation cut, cf. Fig. \ref{qt_lhc_5GeV}.

\subsection{Reliability of NLO calculations with stringent isolation
cuts}\label{stringent}

Let us add one more comment concerning NLO partonic predictions with very 
stringent isolation cuts. In such calculations, the isolation cuts act on the
products of the hard subprocess only. On the other hand, in an actual LHC event, a
cut as severe as $E_{T \, max} = 2.5$~GeV inside a cone $R = 0.3$ or $0.4$
will be nearly saturated by underlying events and pile up.\\

This means that such an isolation cut actually allows almost no transverse
energy deposition from the actual hadronic products of the hard process itself.
This may be most suitable experimentally, and one may think about simulating
such an effect safely in an  NLO partonic calculation by using an effective 
transverse energy cut much more severe than the one experimentally used.
However, requiring that {\it no transverse energy} be deposited in a cone of
fixed size about a photon is {\it not infrared safe}, i.e. it would yield a
divergent result order by order in perturbation theory. This implies that NLO partonic
calculations implemented with finite but very stringent isolation cuts in a
cone of fixed finite size would lead to unreliable results, plagued by
infrared instabilities involving large logarithms of $E_{T \, max}$. What is
more, these infrared nasties would not be located at some isolated point in the
diphoton spectrum (like some elastic boundary or some critical point, as in the
previous subsection), but instead they would extend over its totality, even 
for observables such as the invariant mass distribution. 
The issue of an all order summation of these logarithms of $E_{T \, max}$ 
would have to be investigated in this case.

{\section{Conclusions and perspectives}

We presented an analysis of photon pair production with high invariant mass in 
hadronic collisions, based on a perturbative QCD calculation of full NLO
accuracy. The latter is implemented in the form of a Monte Carlo computer
programme of partonic event generator type, {\it DIPHOX}. The postdictions of 
this study are in reasonable agreement with both WA70 fixed target, and 
preliminary D0 collider data, in the kinematical range where the NLO
approximation is safe, namely away from the elastic boundary of phase space.
Yet more will be learnt from the final analysis of the Tevatron data, and even
more so after the Tevatron run II in the perspective of the LHC. It will then
be worthwhile to perform a more complete phenomenological study.\\

This notwithstanding, there remains room for improvements. A first improvement 
will be to take into account multiple soft gluon effects
in order to calculate infrared sensitive observables correctly. Another 
improvement will concern a more accurate account of contributions beyond NLO,
associated namely with the gluon-gluon initiated subprocess. Among those are 
the NNLO corrections, and even the two loop, so-called double box correction to
$gg \rightarrow \gamma \gamma$, which may be quantitatively important at LHC 
for the background to Higgs search.\\

A better understanding of the effects of isolation, and their interplays with
infrared problems is also required. This concerns the $q_{T}$ distribution near
the critical point $q_{T} = E_{T \, max}$ induced by isolation even
when $E_{T \, max}$ is not small; this concerns also the status of partonic
predictions when $E_{T \, max}$ is chosen very small. Alternatively it would be 
interesting to explore the properties of different isolation criteria, such 
as, for example, the one invented recently by Frixione \cite{frixione}. 
Concerning these last two items, approaches relying on beyond leading order
partonic level calculations, and full event generators like PYTHIA or HERWIG
will be 
complementary.\\

\paragraph{Acknowledgments}
We acknowledge discussions with J. Womersley and T.O. Mentes on the D0 data, J.
Owens on theory vs. data comparisons, and C. Balazs about the theoretical
ingredients inside  the RESBOS code. We thank F. Gianotti, P. Petroff, E.
Richter-Was and V. Tisserand for discussions concerning the Atlas Proposal.
This work was supported in part by the EU Fourth Training  Programme ``Training
and Mobility of Researchers", Network ``Quantum  Chromodynamics and the Deep
Structure of Elementary Particles", contract  FMRX-CT98-0194 (DG 12 - MIHT).

\appendix
\renewcommand{\theequation}{A.\arabic{equation}}
\setcounter{equation}{0}
\section{Technical details on the two photon production}\label{ap}

In this appendix, we give some details on the method used to deal with infrared and soft divergences. For a complete presentation, we refer to \cite{cfg}. The most complicated kinematics happens in the two fragmentation mechanism. Only the two fragmentation contribution will be treated in this appendix, the kinematics of the  other cases can be simply deduced replacing the fragmentation function by a Dirac distribution:
\[
D_{\gamma/k}(x,M_f^2) = \delta(1-x)
\label{eqFRAGMENDELTA}
\]

At the hadronic level, the reaction $H_1(K_1) + H_2(K_2) \rightarrow \gamma(K_3) + \gamma(K_4) + X$  is considered with:
\begin{eqnarray*}
K_1 & = & \frac{\sqrt{S}}{2} \; (1,\, \mbox{\boldmath $0$}, \, 1) \\
K_2 & = & \frac{\sqrt{S}}{2} \; (1,\, \mbox{\boldmath $0$}, \, -1) \\
K_3 & = &  K_{T3} \; (\cosh y_3,\, \mbox{\boldmath $n$}_{3}, \, \sinh y_3) \\
K_4 & = & K_{T4} \; (\cosh y_4,\, \mbox{\boldmath $n$}_{4}, \, \sinh y_4)
\label{eqDEFKINEMATICHADRO}
\end{eqnarray*}
where 
\[
\mbox{\boldmath $n$}_{3}^2 = \mbox{\boldmath $n$}_{4}^2 = 1
\label{eqCARREDEN3N4}
\]
The cross section of the preceding reaction is the sum of the following parts.
\begin{itemize} 
\item[-] The part I (cf. sect. \ref{phase-space-slicing}) contains the infrared, 
         the initial state, and a part of the final state collinear 
	 singularities. Once these divergences have been subtracted, i.e. 
	 cancelled against virtual divergences or absorbed into the bare
	 parton distribution (for the initial state collinear singularities)
	 or the bare fragmentation functions(for the final state collinear 
	 singularities), this part generates three types of finite terms.   
	 \begin{itemize} 
	 \item[(i)] The first type, of infrared origin, has the same kinematics 
	       as the lowest order (LO) terms and is given in \ref{PSCIR} {\it
	       Pseudo cross section for the infrared and virtual parts}. 
	 \item[(ii)] The second type, of initial state collinear origin, has an 
	       extra  integration over the center of mass energy of the hard 
	       scattering, as compared to LO kinematics. For this reason, it is
	       called quasi $2 \rightarrow 2$. It is given in \ref{PSCIC} {\it
	       Pseudo cross sections for the initial state collinear  parts}. 
	 \item[(iii)] There is also a third type, of final state collinear 
	       origin, which involves also an extra integration as compared to
	       LO kinematics. 
	 \end{itemize} 
\item[-] The parts II a and II b contain the rest of the final state collinear
         singularities. Once these divergences have been absorbed into the 
	 bare fragmentation functions, the remaining finite terms
	 involve an extra integration over the relative 
	 momentum of the collinear partons, as compared to LO kinematics. 
	 These terms are combined with those of the so called third type (iii) 
	 above, cf. equations (\ref{eq5PA3}) and (\ref{eq5PA4}). 
	 The resulting contributions are called quasi $2 \rightarrow 2$ as 
	 well. They are given in \ref{PSCFC} {\it Pseudo cross sections for the
	 final state collinear parts}. 
\item[-] The part II c has no divergences. It is given in \ref{PSCF} {\it Cross
	 section for real emission}.
\end{itemize}

\subsection{Cross section for real emission} \label{PSCF}

The cross section is parametrized in the following way:
\begin{eqnarray}
\sigma & = & C_{ij} \; \int dy_3 \; \int dy_4 \; \int dK_{T3} \; \int dK_{T4}
 \nonumber \\
& & \mbox{} \times \left[ \int^1_{x_{3min}} \frac{d x_3}{x_3} \; 
\int_{p_{Tm}}^{p_{T5max}} d p_{T5} \; p_{T5} \raisebox{-2.5mm}{${ {\displaystyle \int \! \! \! \int} \atop {\scriptstyle \Omega_{35}-C_3}}$} 
\; d \phi_{35} \; d y_5  \right. \nonumber \\
& & \mbox{} \times p_{T3} \; D_{\gamma/k}(x_3,M_f^2)
\; \frac{D_{\gamma/l}(x_4,M_f^2)}{p_{T4}} \nonumber \\  
& & \mbox{} \times \frac{F_{i/H_1}(x_1,M^2)}{x_1} \, \frac{F_{j/H_2}(x_2,M^2)}{x_2} \; |M|_3^2 \nonumber \\
& & \mbox{} + \int^1_{x_{4min}} \frac{d x_4}{x_4} \; \int_{p_{Tm}}^{p_{T5max}} 
d p_{T5} \; p_{T5} \raisebox{-2.5mm}{${\displaystyle \int \! \! \! \int} \atop {\scriptstyle \Omega_{45}-C_4}$} \; d \phi_{45} \; d y_5 \nonumber \\
& & \mbox{} \times p_{T4} \; D_{\gamma/l}(x_4,M_f^2) \; \frac{D_{\gamma/k}(x_3,M_f^2)}{p_{T3}} \nonumber \\ 
& & \mbox{} \times \left. \frac{F_{i/H_1}(x_1,M^2)}{x_1} \, \frac{F_{j/H_2}(x_2,M^2)}{x_2} \; |M|_4^2  \right]
\label{eqSIGMAIIC}
\end{eqnarray}
where
\pagebreak
\begin{eqnarray}
x_1 & = & \frac{p_{T3}}{\sqrt{S}} \; e^{- y_3} + \frac{p_{T4}}{\sqrt{S}} \; e^{- y_4} + \frac{p_{T5}}{\sqrt{S}} \; e^{- y_5} \\
& = &\hat{x}_1 + \frac{p_{T5}}{\sqrt{S}} \; e^{- y_5} \nonumber \\
x_2 & = & \frac{p_{T3}}{\sqrt{S}} \; e^{y_3} + \frac{p_{T4}}{\sqrt{S}} \; e^{y_4} + \frac{p_{T5}}{\sqrt{S}} \; e^{y_5} \\
& = & \hat{x}_2 + \frac{p_{T5}}{\sqrt{S}} \; e^{y_5} \nonumber \\
x_{3min} & =&  \frac{2 \, K_{T3}}{\sqrt{S}} \; \cosh y_3
\label{eqDEFX3MIN} \\
x_{4min} & =&  \frac{2 \, K_{T4}}{\sqrt{S}} \; \cosh y_4
\label{eqDEFX4MIN}
\end{eqnarray}
The transverse momenta $p_{T3}$ (resp. $p_{T4}$) are the transverse momenta of the fragmenting partons. They are related to the photon variables by $p_{T3} = K_{T3}/x_3$ (resp. $p_{T4} = K_{T4}/x_4$).
The integration range for the pair of variables $\phi_{35}$ (resp. $\phi_{45}$), $y_5$ is the kinematically allowed range minus a cone in rapidity azimuthal angle $C_3$ (resp. $C_4$) along the $\mbox{\boldmath $p$}_3$ (resp. $\mbox{\boldmath $p$}_4$ ) direction whose size is $R_{th}$. The overall factor $C_{ij}$ reads:
\[
C_{ij} = \frac{\alpha_s^3(\mu^2)}{4 \, S^2 \, \pi \,C_i \, C_j}
\]
and the $C_i$ are given by:
\[
C_i = \left\{ \begin{array}{l}
N \;\; \mbox{for quarks} \\
(N^2-1) \; \; \mbox{for gluons}
\end{array} \right. 
\]
\\ 

The matrix element squared \footnote{An overall factor of the matrix element squared containing the average on spins and colors of the initial state and the coupling constant has been put into the coefficient $C_{ij}$}, taken from the first reference of \cite{abfs} and \cite{ellis-sexton}, has been split into two parts:
\[
|M|^2 = |M|^2_3 + |M|^2_4
\]
The first part $|M|^2_3$ contains final state collinear singularities arising
when $\mbox{\boldmath $p$}_3$ // $\mbox{\boldmath $p$}_5$ and the second part $|M|^2_4$ contains final
state collinear singularities arising when $\mbox{\boldmath $p$}_4$ // $\mbox{\boldmath $p$}_5$. More
precisely, the matrix element squared can be written as a weighted sum of
eikonal factors $E_{ab}$ plus a term free of infrared or collinear
singularities:
\begin{equation}
|M|^2_{ij \rightarrow klm} = 
\sum_{a=1}^4 \sum_{b=a+1}^4 H_{ab}(p_5) E_{ab}
+ G(p_5)
\label{eqM2IJKLM}
\end{equation}
where
\[
E_{ab}  = \frac{p_a.p_b}{p_a.p_5 \; p_b.p_5}
\]
Using:
\begin{equation}
\frac{1}{p_3.p_5 \; p_4.p_5} = \frac{1}{p_1.p_5 + p_2.p_5} 
\left(
\frac{1}{p_3.p_5} + \frac{1}{p_4.p_5} 
\right)
\end{equation}
we get:
\pagebreak
\begin{eqnarray*}
|M|^2_3 & = & \frac{1}{2} H_{12}(p_5) E_{12} + H_{13}(p_5) E_{13}
+ H_{23}(p_5) E_{23} + H_{34}(p_5) E'_{34} + \frac{1}{2} G(p_5) \\
|M|^2_4 & = & \frac{1}{2} H_{12}(p_5) E_{12} + H_{14}(p_5) E_{14}
+ H_{24}(p_5) E_{24} + H_{34}(p_5) E''_{34} + \frac{1}{2} G(p_5)
\end{eqnarray*}
with
\begin{eqnarray*}
E'_{34} & = & \frac{p_3.p_4}{p_1.p_5 + p_2.p_5} \; \frac{1}{p_3.p_5} \\
E''_{34} & = & \frac{p_3.p_4}{p_1.p_5 + p_2.p_5} \; \frac{1}{p_4.p_5}
\end{eqnarray*}
In order that the infrared divergences cancel, and the collinear singularities factorize out, the coefficients $H_{ab}$ have to fulfill:
\begin{eqnarray*}
\lefteqn{\frac{C_i}{C_{i'}} \, a_{i'i}^{(d)}(z_1) \, |M|^{2B}_{i'j \rightarrow kl} =} \\
& & z_1 \; \biggl[ H_{12} \Bigl( (1-z_1) \, p_1 \Bigr) + H_{13} \Bigl( (1-z_1) \, p_1 \Bigr) 
+ H_{14} \Bigl( (1-z_1) \, p_1 \Bigr) \biggr] \\
\lefteqn{\frac{C_j}{C_{j'}} \, a_{j'j}^{(d)}(z_2) \, 
|M|^{2B}_{ij' \rightarrow kl} =} \\
& &  z_2 \; \biggl[ H_{12} \Bigl( (1-z_2) \, p_2 \Bigr)
+ H_{23} \Bigl( (1-z_2) \, p_2 \Bigr) + H_{24} \Bigl( (1-z_2) \, p_2 \Bigr) 
\biggr] \\  
\lefteqn{a_{kk'}^{(d)}(z_3) \, |M|^{2B}_{ij \rightarrow k'l} =} \\ 
& & z_3 \;
\Biggl[H_{13} \left( \frac{1-z_3}{z_3} \, p_3 \right)
+ H_{23} \left( \frac{1-z_3}{z_3} \, p_3 \right) + H_{34}
\left( \frac{1-z_3}{z_3} \, p_3 \right)
\Biggr] \\ 
\lefteqn{a_{ll'}^{(n)}(z_4) \, |M|^{2B}_{ij \rightarrow kl'}  =} \\
& & z_4 \; 
\Biggl[H_{14} \left( \frac{1-z_4}{z_4} \, p_4 \right)
+ H_{24} \left(\frac{1-z_4}{z_4} \, p_4 \right) + H_{34}
\left( \frac{1-z_4}{z_4} \, p_4 \right) 
\Biggr] \\
\end{eqnarray*}
In particular, the cancellation of infrared divergences is insured by:
\begin{eqnarray*}
H_{12}(0) + H_{13}(0) + H_{14}(0) & = & a^{(d)}_{ii}(1)
|M|^{2B}_{ij \rightarrow kl} \\ 
H_{12}(0) + H_{23}(0) + H_{24}(0) & = & a^{(d)}_{jj}(1)
|M|^{2B}_{ij \rightarrow kl} \\ 
H_{13}(0) + H_{23}(0) + H_{34}(0) & = & a^{(d)}_{kk}(1)
|M|^{2B}_{ij \rightarrow kl} \\ 
H_{14}(0) + H_{24}(0) + H_{34}(0) & = & a^{(d)}_{ll}(1)
|M|^{2B}_{ij \rightarrow kl} .
\end{eqnarray*}
The functions $a^{(d)}_{ij}(z)$ will be given in equation (\ref{eqAP}).\\

In equation~(\ref{eqSIGMAIIC}), the integration domain for the rapidities and
the transverse momenta of the two photons is in general limited by
experiments. The integration over $p_{T5}$ is constrained by:
\[
p_{T5}^2 < S \, (1-\hat{x}_1) \, (1-\hat{x}_2).
\]

\subsection{Pseudo cross sections for the initial state collinear parts}\label{PSCIC}

The finite part associated to the collinear divergence $\mbox{\boldmath $p$}_1$ // $\mbox{\boldmath $p$}_5$ is given by: 
\begin{eqnarray}
\sigma_{5//1} & = & \int dy_3 \; \int dy_4 \; \int dK_{T3} \; \int dK_{T4} \; \int^1_{x_{3 min}} \frac{dx_3}{x_3} \; \int^1_{x_{4 min}}
\frac{dx_4}{x_4} \nonumber \\
& & \mbox{} \times \frac{\alpha_s(\mu^2)}{2 \, \pi} \; 
C_{ij}^B  \;  p_{T} \; \delta{(p_{T3}-p_{T4})} D_{\gamma/k}(x_3,M^2_f) \; D_{\gamma/l}(x_4,M^2_f) \nonumber \\
& & \mbox{} \times 
\int^1_{x_1^0} \frac{dz_1}{z_1} \;
 \frac{F_{i/H_1} \left(\frac{x_1^0}{z_1},M^2 \right)}{x_1^0} \;
\frac{F_{j/H_2}(x_2^0,M^2)}{x_2^0} \; |M|^{2B}_{i'j \rightarrow kl} \nonumber \\
& & \mbox{} \times
\frac{C_i}{C_{i'}} \; \Biggl[
\frac{a^{(d-4)}_{i'i}(z_1)}{(1-z_1)_+} + \ln 
\biggl( \frac{p^2_{Tm}}{M^2} \biggr) \,
P^{(4)}_{i'i}(z_1) -f_{i'i}(z_1) 
\Biggr]
\label{eqSIGMA5//1}
\end{eqnarray}
where the variables $x_1^0$ (resp. $x_2^0$) are defined by:
\begin{eqnarray*}
x_1^0 & = & \frac{p_{T}}{\sqrt{S}} \; \left( e^{- y_3} + e^{- y_4} \right)
\\ 
x_2^0 & = & \frac{p_{T}}{\sqrt{S}} \; \left( e^{y_3} + e^{y_4} \right)
\end{eqnarray*}
and $p_T$ stands for $p_{T3}$ or $p_{T4}$.\\

The finite part associated to the collinear divergence $\mbox{\boldmath $p$}_2$ // $\mbox{\boldmath $p$}_5$ is given by: 
\begin{eqnarray}
\sigma_{5//2} & = & \int dy_3 \; \int dy_4 \; \int dK_{T3} \; \int dK_{T4} \; \int^1_{x_{3 min}} \frac{dx_3}{x_3} \; \int^1_{x_{4 min}} \frac{dx_4}{x_4} \nonumber \\
& & \mbox{} \times \frac{\alpha_s(\mu^2)}{2 \, \pi}
C_{ij}^B  \; p_{T} \; \delta{(p_{T3}-p_{T4})} \; D_{\gamma/k}(x_3,M^2_f) \; D_{\gamma/l}(x_4,M^2_f) \nonumber \\
& & \mbox{} \times 
\int^1_{x_2^0} \frac{dz_2^0}{z_2} \;
\frac{F_{i/H_1}(x_1^0,M^2)}{x_1^0} \;
\frac{F_{j/H_2} \left( \frac{x_2^0}{z_2},M^2 \right)}{x_2^0} \; |M|^{2B}_{ij' \rightarrow kl} \nonumber \\
& & \mbox{} \times \frac{C_j}{C_{j'}} \;
\Biggl[ \frac{a^{(d-4)}_{j'j}(z_2)}{(1-z_2)_+} + \ln 
\biggl( \frac{p^2_{Tm}}{M^2} \biggr) \,
P^{(4)}_{j'j}(z_2) -f_{j'j}(z_2^0) 
\Biggr]
\label{eqSIGMA5//2}
\end{eqnarray}
with
\[
C^{B}_{ij} = \frac{2 \, \pi \alpha_s^2(\mu^2)}{4 \, S^2 \, C_i \, C_j}
\]

The functions $a^{(d-4)}_{ij}(z)$, $P^{(4)}_{ij}(z)$ and $f_{ij}(z)$ will be
defined at the end of this appendix cf. equations from (\ref{eqANM4GG}) to (\ref{eqANM4QG}), (\ref{eqAP}) and (\ref{eqFFUNCTION}).

\subsection{Pseudo cross section for the final state collinear parts}\label{PSCFC}

These parts contain the collinear singularities which have been absorbed into the bare fragmentation functions. \\

The finite part associated to the collinear divergence $\mbox{\boldmath $p$}_3$ // $\mbox{\boldmath $p$}_5$ is given by:
\begin{eqnarray}
\sigma_{5//3} & = & \int dy_3 \; \int dy_4 \; \int dK_{T3} \; \int dK_{T4} \;  \int^1_{x_{3 min}} \frac{dx_3}{x_3} \; \int^1_{z_{3 min}} \frac{dz_3}{z_3}  \nonumber \\
& & \mbox{} \times \frac{\alpha_s(\mu^2)}{2 \, \pi} \; C_{ij}^B \; 
D_{\gamma/k}(x_3,M^2_f) \; D_{\gamma/l}(x'_4,M^2_f) \nonumber \\
& & \mbox{} \times 
\frac{F_{i/H_1}(x'_1,M^2)}{x'_1} \; 
\frac{F_{j/H_2}(x'_2,M^2)}{x'_2} \; |M|^{2B}_{ij \rightarrow k'l}
\nonumber \\
& & \mbox{} \times \Biggl[ 
\frac{a^{(d-4)}_{kk'}(z_3)}{(1-z_3)_+} 
+
\ln \biggl( \frac{p^2_{T3}}{M^2_f} \biggr) \,
P^{(4)}_{kk'}(z_3) - d_{kk'}(z_3) \label{eq5PA3} \\
& & \mbox{} + 2 \,
\Biggl( \frac{ \ln (1-z_3)}{(1-z_3)} \Biggr)_+ a^{(4)}_{kk'}
(z_3) +
\ln(R^2)  \, \frac{a^{(4)}_{kk'}(z_3)}{(1-z_3)} \, \Theta(z_{3 m} -
z_3) 
\Biggr] \nonumber
\end{eqnarray}
whereas the finite part associated to the collinear divergence $\mbox{\boldmath $p$}_4$ // $\mbox{\boldmath $p$}_5$ is given by:
\begin{eqnarray}
\sigma_{5//4} & = & \int dy_3 \; \int dy_4 \; \int dK_{T3} \; \int dK_{T4} \;  \int^1_{x_{4 min}} \frac{dx_4}{x_4} \;
 \int^1_{z_{4 min}} \frac{dz_4}{z_4} \nonumber \\
& & \mbox{} \times \frac{\alpha_s(\mu^2)}{2 \, \pi} \; C_{ij}^B \;
D_{\gamma/k}(x''_3,M^2_f) \; D_{\gamma/l}(x_4,M^2_f)  \nonumber \\
& & \mbox{} \times 
\frac{F_{i/H_1}(x''_1,M^2)}{x''_1} \;
\frac{F_{j/H_2}(x''_2,M^2)}{x''_2} \; |M|^{2B}_{ij \rightarrow kl'}
\nonumber \\
& & \mbox{} \times \Biggl[ 
\frac{a^{(d-4)}_{ll'}(z_4)}{(1-z_4)_+} +
\ln 
\biggl( \frac{p^2_{T4}}{M^2_f} \biggr)
P^{(4)}_{ll'}(z_4) -d_{ll'}(z_4) \label{eq5PA4} \\  
& & \mbox{} + 2 
\Biggl( \frac{ \ln (1-z_4)}{(1-z_4)} \Biggr)_+ a^{(4)}_{ll'}
(z_4) + \ln(R^2) \, \frac{a^{(4)}_{ll'}(z_4)}{(1-z_4)} \, \Theta(z_{4 m} -
z_4)
\Biggr] \nonumber
\end{eqnarray}

The functions $a^{(4)}_{ij}(z)$ and $d_{ij}(z)$ will be also defined at the end of the appendix cf. equations from (\ref{eqA4GG}) to (\ref{eqA4QG}) and (\ref{eqDFUNCTION}). The variables $z_{3m}$, $z_{4m}$, $x'_1$, $x'_2$, $x''_1$ and $x''_2$ appearing in equations (\ref{eq5PA3}) and (\ref{eq5PA4}) are given by:
\begin{eqnarray*}
z_{3m} & = & \frac{p_{T3}}{p_{T3} + p_{Tm}} \\
z_{4m} & = & \frac{p_{T4}}{p_{T4} + p_{Tm}} \\
x'_1 & = & \frac{p_{T3}+p_{T5}}{\sqrt{S}} \; \left( e^{y_3} + e^{y_4} \right) \\
x'_2 & = & \frac{p_{T3}+p_{T5}}{\sqrt{S}} \; \left( e^{-y_3} + e^{-y_4} \right) \\
x''_1 & = & \frac{p_{T4}+p_{T5}}{\sqrt{S}} \; \left( e^{y_3} + e^{y_4} \right) \\
x''_2 & = & \frac{p_{T4}+p_{T5}}{\sqrt{S}} \; \left( e^{-y_3} + e^{-y_4} \right)
\end{eqnarray*}

\subsection{Pseudo cross section for the infrared and virtual parts}\label{PSCIR}

This pseudo cross section is given by:
\begin{eqnarray}
\sigma_{ir} & = & \int dy_3 \; \int dy_4 \; \int dK_{T3} \; \int dK_{T4} \nonumber \\
& & \mbox{} \times \frac{\alpha_s(\mu^2)}{2 \; \pi} \;
 C_{ij}^B \;
\int^1_{x_{3 min}} \frac{dx_3}{x_3} \; \int^1_{x_{4 min}}
\frac{dx_4}{x_4} \; p_{T} \; \delta{(p_{T3}-p_{T4})}
\nonumber \\
& & \mbox{} \times 
D_{\gamma/k}(x_3,M^2_f) \; D_{\gamma/l}(x_4,M^2_f) \;
\frac{F_{i/H_1}(x_1^0,M^2)}{x_1^0} \;
\frac{F_{j/H_2}(x_2^0,M^2)}{x_2^0} \; \nonumber \\
& & \mbox{} \times
\Biggl\{ - \Biggl[ 
\ln \biggl( \frac{p_T^2}{S} \biggr) (b_{kk} + b_{ll}) + 
\ln \biggl( \frac{p_{Tm}^2}{S} \biggr) (b_{ii} + b_{jj}) 
\Biggr] 
|M|^{2B}_{ij \rightarrow kl} \nonumber \\
& & \mbox{} + \ln \biggl( \frac{p_{Tm}^2}{S} \biggr) 
\Biggl(
\sum_{i<j} H_{ij}(0) \ln \biggl( \frac{2 p^0_i.p^0_j}{S} \biggr) 
\Biggr) \nonumber \\ 
& & \mbox{} - \frac{1}{2} \ln 
\biggl( \frac{p_{T}^2}{S} \biggr)
\ln \biggl( \frac{p_{Tm}^2}{S} \biggr) 
\Bigl[ H_{13}(0) + H_{14}(0) + H_{23}(0) + H_{24}(0) + 2 H_{34}(0) \Bigr]
\nonumber \\
& & \mbox{} + \frac{1}{4}\ln^2 \biggl( \frac{p_{T}^2}{S} \biggr)
\Bigl[ H_{13}(0) + H_{14}(0) + H_{23}(0) + H_{24}(0) + 2 H_{34}(0) \Bigr]
\nonumber \\
& & \mbox{} - \frac{1}{4}\ln^2 \biggl( \frac{p_{Tm}^2}{S} \biggr)
\Bigl[ 2 H_{12}(0) + H_{13}(0) + H_{14}(0) + H_{23}(0) + H_{24}(0) \Bigr]
\nonumber \\
& & \mbox{} + \frac{H_{34}(0)}{\pi} 
\Bigl[ A_{34}(y^{\star}) + A_{34} (-y^{\star}) \Bigr]  + F(\hat{s},\hat{t},\hat{u})
\Biggr\} 
\label{eqIRTYPE}
\end{eqnarray}
with $p_T = p_{T3} = p_{T4}$. The terms $b_{ii}$ are defined in equations (\ref{eqBGG}) and (\ref{eqBQQ}).
In the equation (\ref{eqIRTYPE}), $y^{\star} = (y_3-y_4)/2$ and the function $A(x)$
is given by:
\begin{eqnarray*}
A(x) & = & \pi \, \ln(2) \, \ln( 4 \, \cosh^2(y^{\star})) + 2 \, y^{\star} \, \sinh(2 \, y^{\star}) \, \int^{\pi}_0 d \phi \; \frac{\ln ( \sin \phi)}{\cosh(2 \, x) + \cos (2 \, \phi)}  \nonumber \\
& & \mbox{} + 4 \; \int^{\pi}_0 d \phi \; \frac{\sin (2 \, \phi)}{\cosh(2 \, x) + \cos (2 \, \phi)} \; \ln(\sin \phi) \; \arctan \left( \frac{\sin \phi}{1 - \cos \phi} \right)
\end{eqnarray*}
The function $F$ is the finite part of the virtual term and the variables
$\hat{s}$, $\hat{t}$ and $\hat{u}$ are the Mandelstam variables of the $2 \rightarrow 2$ processes:
\begin{eqnarray*}
\hat{s} = (p_1^0+p_2^0)^2 \\
\hat{t} = (p_1^0-p_3^0)^2 \\
\hat{u} = (p_2^0-p_3^0)^2
\end{eqnarray*}
where the 4-vectors $p_i^{0}$ are the infrared limits of the 4-vectors $p_i$.

\subsection{Altarelli-Parisi Kernels}

We will give in this appendix the expressions of the functions
$a_{ij}(z)$ and $b_{ij}$. These functions are defined by:
\pagebreak
\begin{eqnarray}
P^{(d)}_{ij}(z) & = & \frac{a^{(d)}_{ij}(z)}{(1-z)_+} + b_{ij} \, \delta(1-z)
 \nonumber \\
& = & \frac{a^{(4)}_{ij}(z) - \epsilon \, a^{(d-4)}_{ij}(z)}{(1-z)_+}
+ b_{ij}  \, \delta(1-z) \nonumber \\
& = & P^{(4)}_{ij}(z) - \epsilon \; \frac{a^{(d-4)}_{ij}(z)}{(1-z)_+}
\label{eqAP}
\end{eqnarray}
where $P^{(4)}_{ij}$ (resp. $P^{(d)}_{ij}$) are the Altarelli-Parisi Kernels in four (resp. $d$) dimensions. So the expressions for the functions $a^{(4)}_{ij}(z)$, $a^{(d-4)}_{ij}(z)$ and $b_{ij}$ are given by:
\begin{eqnarray}
a^{(4)}_{gg}(z) & = & 2 \, N \, 
\left( z + \frac{(1-z)^2}{z} + z (1-z)^2 \right)
\label{eqA4GG} \\
a^{(4)}_{qq}(z) & = & C_F \, (1+z^2)
\label{eqA4QQ} \\
a^{(4)}_{gq}(z) & = & C_F \, 
\left( \frac{1+(1-z)^2}{z} \right) (1-z)
\label{eqA4GQ} \\
a^{(4)}_{qg}(z) & = & T_F \, (z^2 + (1-z)^2 ) \, (1-z)  
\label{eqA4QG}
\end{eqnarray}
where $N$ is the number of colors, $C_F = (N^2-1)/(2 N)$ and $T_F =
1/2$. The extra part needed to get the functions $a$ in $d$ dimensions
($a^{(d)}_{ij}(z) = a^{(4)}_{ij}(z) - \epsilon a^{(d-4)}_{ij}(z)$) is
given by:
\begin{eqnarray}
a^{(d-4)}_{gg}(z) & = & 0
\label{eqANM4GG} \\
a^{(d-4)}_{qq}(z) & = & C_F \, (1-z)^2
\label{eqANM4QQ} \\
a^{(d-4)}_{gq}(z) & = & C_F \, z \, (1-z)
\label{eqANM4GQ} \\
a^{(d-4)}_{qg}(z) & = & 2 \, T_F \, z \, (1-z)^2 
\label{eqANM4QG}
\end{eqnarray}
The coefficients $b_{ij}$ read:
\begin{eqnarray}
b_{gg} & = & \frac{(11 \, N - 2 \, N_F)}{6}
\label{eqBGG} \\
b_{qq} & = & \frac{3}{2} \, C_F
\label{eqBQQ}
\end{eqnarray}

The function $f_{ij}(x)$ and $d_{ij}(z)$ define the factorisation scheme for respectively initial state and final state collinear singularities. In the $\overline{MS}$ scheme, we have:
\begin{eqnarray}
f_{ij}(z) = 0
\label{eqFFUNCTION} \\
d_{ij}(z) = 0 
\label{eqDFUNCTION}
\end{eqnarray}

\renewcommand{\theequation}{B.\arabic{equation}}
\setcounter{equation}{0}
\section{Cancellation of the $p_{Tm}$ and $R_{th}$ dependences}\label{appendix-b}

In this appendix, we give further details on the cancellation of the $p_{Tm}$ 
and $R_{th}$ dependences in observables calculated according to the method used
in this article. \\

In the conical parts II a and II b, the $d$-dimensional integration over 
particle 5 in $C_{i}$, $i=3,4$, reads schematically:
\begin{eqnarray}\label{eqQW1}
\sigma_{i} & = & \int_{p_{Tm}}^{p_{T5max}} d p_{T5} \; p_{T5}^{-1-2\, \epsilon} 
\nonumber \\
&& \;\;\;\;\;\; \int_{C_{i}} d \phi_{i5} \; d y_5 \; 
\sin^{-\, 2\, \epsilon} \, \phi_{i5}  \; 
\frac{F(p_{T5},\phi_{i5},y_5)}{\cosh(y_i-y_5)-\cos \phi_{i5}}
\end{eqnarray}
The term generating the final state collinear pole 
($\mbox{\boldmath $p$}_{5}$//$\mbox{\boldmath $p$}_{i}$) has been explicitly
written, and the remaining quantity $F(p_{T5},\phi_{i5},y_5)$ is a regular
function. In the parts II a and II b, the same subtraction method as in 
\cite{cfg} is used, and the following contribution is added and subtracted:
\begin{equation}\label{eqQW2}
\sigma^{sub}_{i} = \int_{p_{Tm}}^{p_{T5max}} d p_{T5} \; 
p_{T5}^{-1-2\, \epsilon} 
\int_{C_{i}} d \phi_{i5} \; d y_5 \; \phi_{i5}^{-\, 2\, \epsilon}  \; 
\frac{2 \, F(p_{T5},0,y_i)}{(y_i-y_5)^2+\phi_{i5}^2}
\end{equation}
In the cylindrical part I, the finite terms produced by the integration over
particle 5 are approximated: all the terms depending logarithmically on 
$p_{Tm}$ are kept, whereas terms proportional to powers of $p_{Tm}$
are neglected. Notice that this differs from the subtraction method implemented 
in the cylinder in \cite{cfg}, which kept the exact $p_{Tm}$ dependence. \\

In summary, the present method is an admixture of the phase space slicing and
subtraction methods, at variance with what has been done in \cite{cfg}. It 
ensures the exact cancellation of the unphysical parameter $R_{th}$ dependence
between part II c and parts II a, II b whereas only an approximated
cancellation of the unphysical parameter $p_{Tm}$ dependence between parts II
c, II a and  II b and part I occurs. \\

We checked carefully that the dependences on the unphysical parameters drop out.
This point is illustrated  by the $p_{Tm}$ dependence (at fixed
$R_{th}=0.1$) and the $R_{th}$ dependence (at fixed $p_{Tm}= 0.1$ GeV), of the
higher order (HO) part of integrated cross section (the lowest order (LO) part being independent of these parameters) 
\[
\sigma^{HO} = \int_{m_{min}}^{m_{max}} d \, m_{\gamma \gamma} \; 
\frac{d \, \sigma^{HO}}{d \, m_{\gamma \gamma}}
\]
shown on Figs. \ref{ptm_dep_dir} and \ref{ptm_dep_frag}.  We display
separately the $q \bar{q}$ and $q g$ initiated contributions to the ``direct"
on Fig. \ref{ptm_dep_dir}, and the ``one-" and ``two fragmentation"
mechanisms, on Fig. \ref{ptm_dep_frag}. To be definite, the integration bounds
are taken to be $m_{min} = 80$ GeV,  $m_{max} = 1500$ GeV, the cuts $p_{T3}$,
$p_{T4} \geq 25$ GeV, $|y_{3,4}| \leq 2.5$ are applied, and the MRST2 set of
parton distribution functions with the scale choice  $M = \mu = M_f = m_{\gamma
\gamma}/2$ are used; let us emphasize however that the pattern obtained does
not depend on these details.\\
 
The quantity $\sigma^{HO}$ does not depend on $R_{th}$ and, in principle, it becomes
independent of $p_{Tm}$ at small enough $p_{Tm}$. 
To show these features more clearly, the observable displayed is the ratio $R_{m_{\gamma \gamma}}$  
defined as follows:
\begin{equation}\label{eqDEFRMGG}
R_{m_{\gamma \gamma}} =  
\frac {1}{A} \int_{m_{min}}^{m_{max}} 
d \, m_{\gamma \gamma} \; \frac{d \, \sigma^{HO}}{d \, m_{\gamma \gamma}}
\end{equation}
The integrated  cross section is normalized to be asymptotically 1 in order
to show the size of the relative error bars. However taking the denominator $A$ equal to the calculated $\sigma^{HO}$
for the smallest value of $p_{Tm}$ may be numerically unsuitable. Indeed, when $p_{Tm}$ becomes
smaller and smaller, numerical cancellations between larger and larger
contributions occur and the error bars coming from the Monte Carlo integration
become larger and larger. These numerical fluctuations affect the behavior in
the limit $p_{Tm} \rightarrow 0$. In order to bypass these technical complications, $A$ is taken to be the averaged value
of those of the integrated cross sections $d \, \sigma^{HO} / d \, m_{\gamma
\gamma}$ which are  consistent with each other within the error bars.
For
instance, for the $p_{Tm}$ dependence of the ``direct" contribution, the 
average is taken over the values corresponding to the three smallest $p_{Tm}$
because the fourth one is not consistent with the others in the error bars. In
addition, in the case of the direct contribution, the two partonic reactions $q
\bar{q}$ and $q g$ have been split because, for the above choices of scales,
the two integrated contributions are large and of opposite signs. As expected,
$R_{m_{\gamma \gamma}}$ does not depend on $R_{th}$ and approaches 1 as $p_{Tm}
\rightarrow 0$. 
Let us notice that one can wonder whether large relative fluctuations do not appear again when the two contributions of the ``direct" are added. Indeed, the relative fluctuations of the HO terms are larger for the sum than for each parts, but these HO terms are small compared to the LO part ($\sigma^{HO} \sim {\mathcal O}( 1 \%) \, \sigma^{LO}$) hence the ``physical" cross section (LO+HO) is sufficiently stable.
When the parameter $p_{Tm}$ is chosen small enough with respect
to $p_{T3}$ and  $p_{T4}$, the neglected terms power behaved in $p_{Tm}$ can be
safely dropped out. In practice, we observe that  $p_{Tm}$ values of the order
of half a percent of the minimum $p_{T3}$ and $p_{T4}$, i.e. $p_{Tm} \leq 0.1$
GeV, fulfill these requirements.
Before embarking in a long phenomenological study, the user of the DIPHOX code is advised to check whether the value of the parameter $p_{Tm}$ to be used is small enough to neglect safely the power corrections of $p_{Tm}$.

\begin{figure}[htb]
\begin{center}
\mbox{\epsfig{file=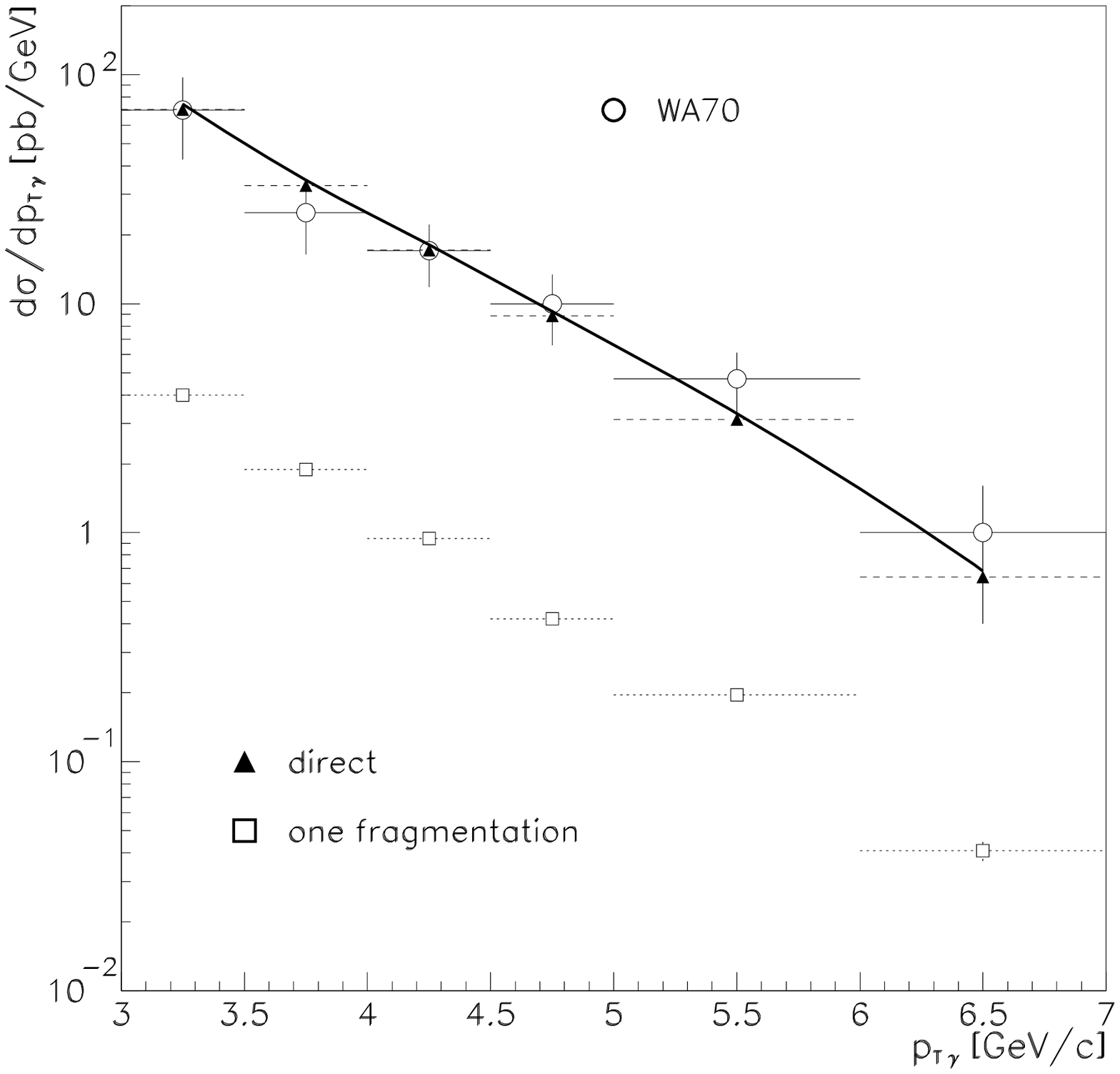,height=14cm}}
\end{center}
\caption{
 Diphoton differential cross section $d\sigma/dp_T$ vs. $p_T$, 
 the transverse energy of each photon, in $\pi^{-}$-proton collisions at
 $\sqrt{S}= 22.9$ GeV.
 Data points from the WA70 collaboration ~\protect\cite{wa70cross}.
 The solid line is the full contribution
 with scales $M=\mu=M_f=0.275 \, (p_T (\gamma_1) + p_T (\gamma_2))$.}
\label{flc}
\end{figure}

\begin{figure}[htb]
\begin{center}
\mbox{\epsfig{file=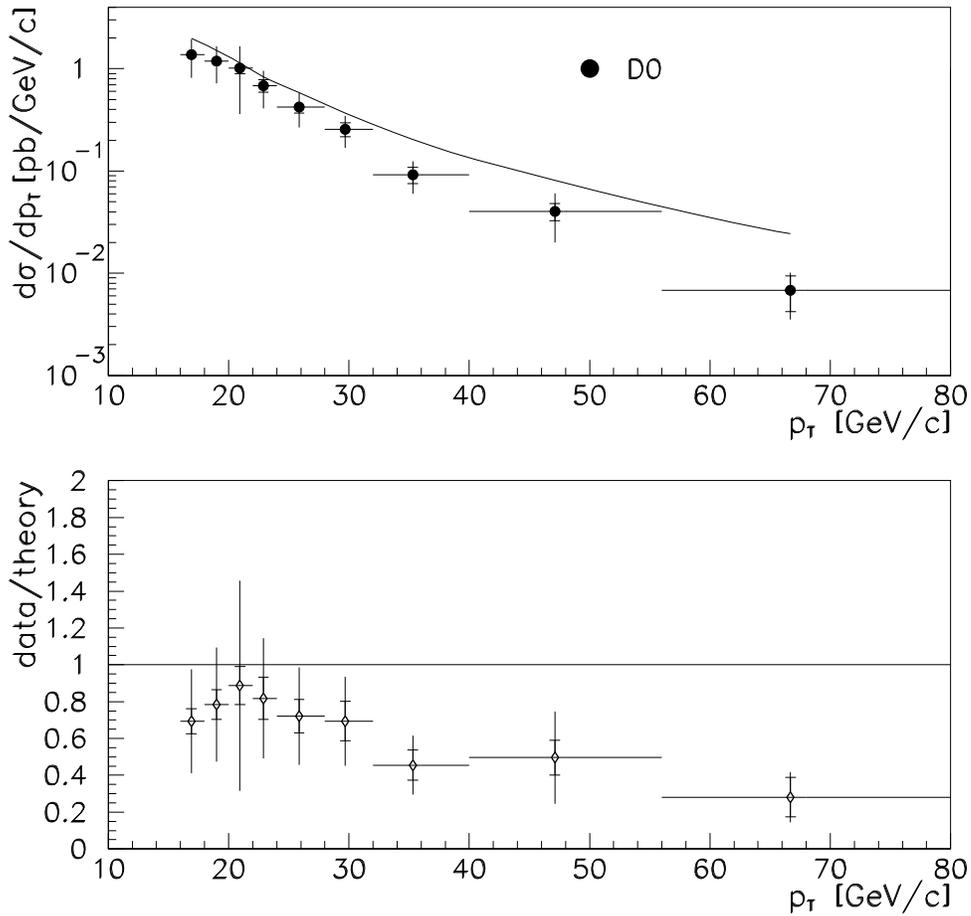,height=14cm}}
\end{center}
\caption{
 Diphoton differential cross section $d\sigma/dp_T$ vs $p_T$, 
 the transverse energy of each photon, at Tevatron, $\sqrt{S}=1.8$ TeV.
  Preliminary data points (statistical errors and systematics in quadrature) 
 from the D0 collaboration~\protect\cite{d0} are compared to the theoretical
 predictions: the full NLO prediction is shown as
 the solid line. The ratio data/(full NLO theory)
 is shown below.
 }
\label{flc1}
\end{figure}

\begin{figure}[htb]
\begin{center}
\mbox{\epsfig{file=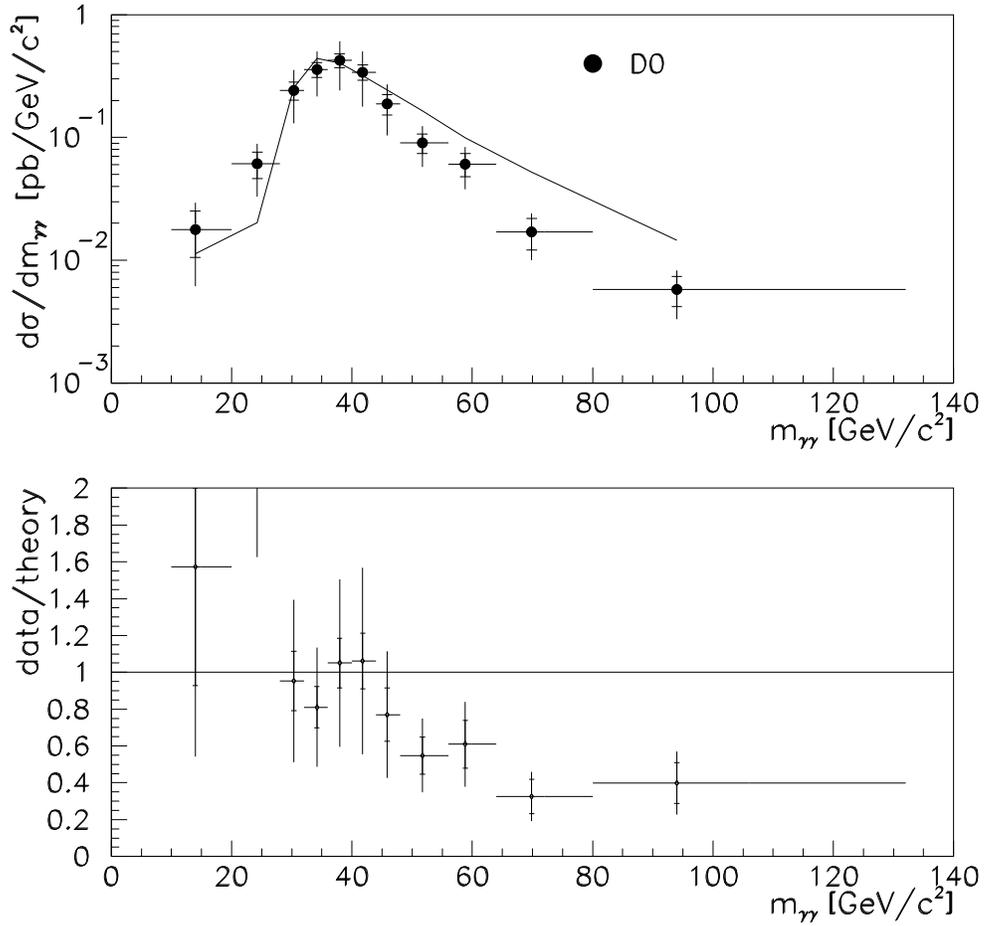,height=14cm}}
\end{center}
\caption{
 Diphoton differential cross section $d\sigma/dm_{\gamma \gamma}$ 
 vs. $m_{\gamma\gamma}$, the mass of the photon pair, at Tevatron,
 $\sqrt{S}=1.8$ TeV.
 Preliminary data points (statistical errors and systematics in quadrature)
 from the D0 collaboration~\protect\cite{d0} are compared to the theoretical
 predictions: the full NLO prediction is shown as the solid line.
 }
\label{flc2}
\end{figure}

\begin{figure}[htb]
\begin{center}
\mbox{\epsfig{file=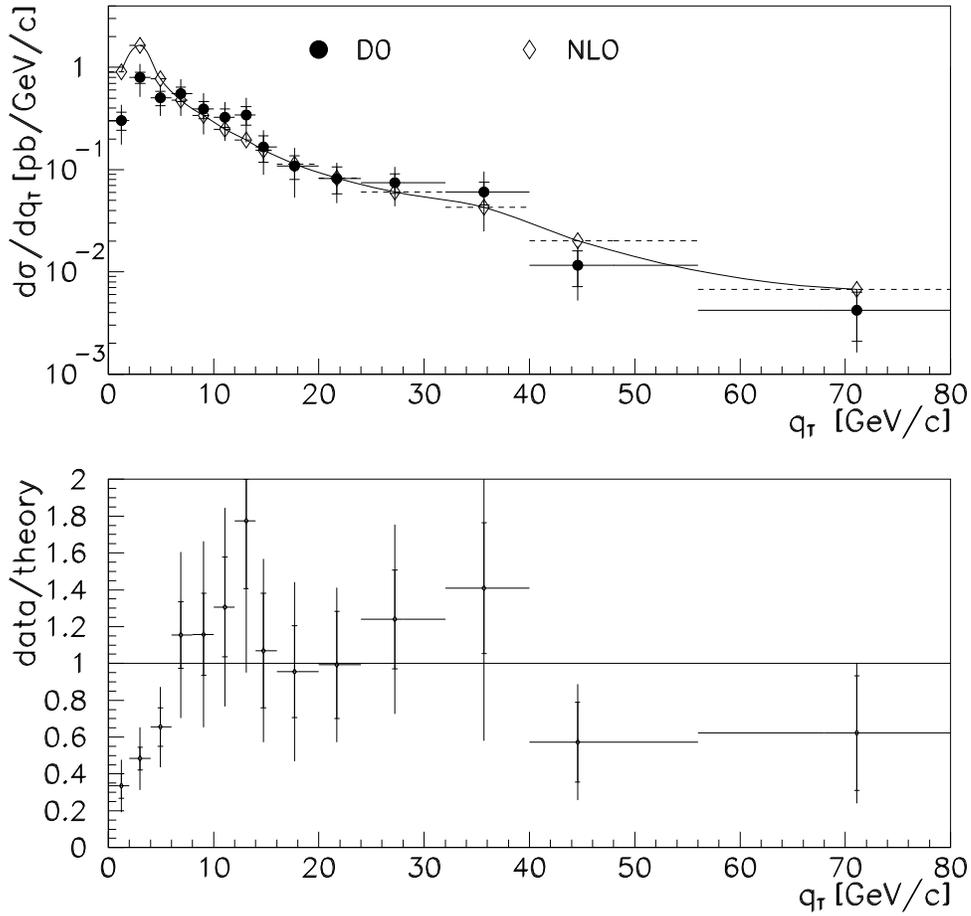,height=14cm}}
\end{center}
\caption{
 Diphoton differential cross section $d\sigma/dq_T$ 
 vs. $q_T$, the transverse momentum  of the photon pair, at 
 Tevatron, $\sqrt{S}=1.8$ TeV.
 Preliminary data points (statistical errors and systematics in quadrature) 
 from the D0 collaboration~\protect\cite{d0} are compared to the theoretical
 predictions: the full NLO prediction is shown as the solid line
  }
\label{flc3}
\end{figure}

\begin{figure}[htb]
\begin{center}
\mbox{\epsfig{file=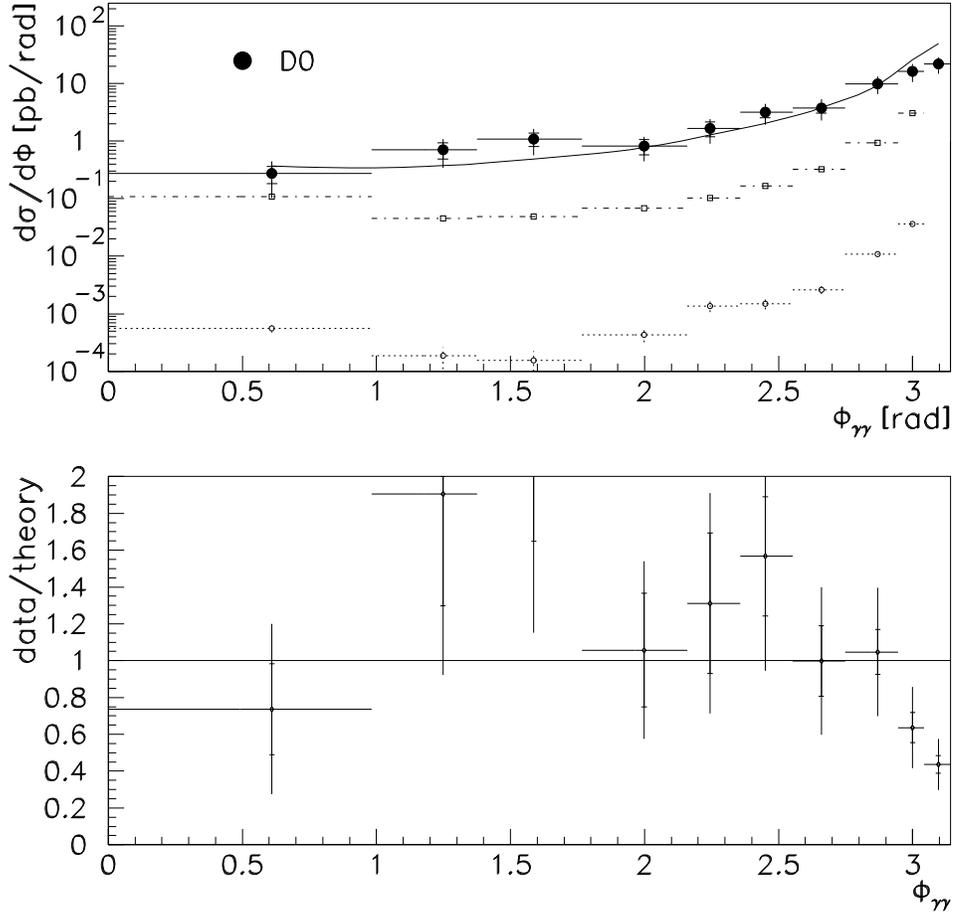,height=14cm}}
\end{center}
\caption{
Diphoton differential cross section $d\sigma/d \phi_{\gamma\gamma}$  
vs. $\phi_{\gamma\gamma}$, the azimuthal angle between the two 
photons, at Tevatron, $\sqrt{S}=1.8$ TeV.
 Preliminary data points (statistical errors and systematics in quadrature) 
 from the D0 collaboration~\protect\cite{d0} are compared to the theoretical
 predictions: the full NLO prediction is shown as the solid line
 while open squares (open circles) represent the single 
 (double) fragmentation contribution.}
\label{flc4}
\end{figure}
\begin{figure}[htb]
\begin{center}
\mbox{\epsfig{file=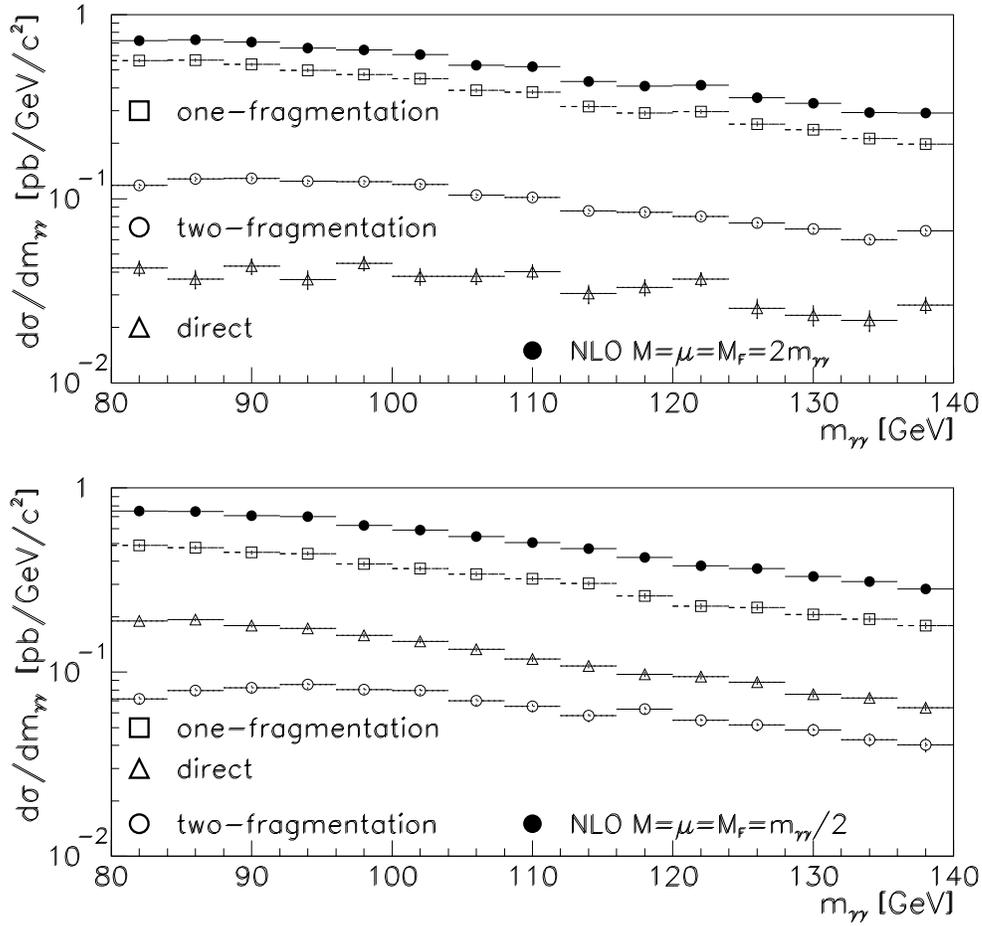,height=14cm}}
\end{center}
\caption{
 Splitting of the diphoton differential cross section 
 $d\sigma/dm_{\gamma \gamma}$ at LHC,
 $\sqrt{S}=14$ TeV without isolation, into the ``direct",``one
 fragmentation" and ``two fragmentation" components, shown for two different
 choices of scales. The following kinematic cuts are applied:
 $p_{T}(\gamma_{1}) \geq 40$ GeV, $p_{T}(\gamma_{2}) \geq 25$ GeV, 
 $|y(\gamma_{1,2})| \leq 2.5$. 
 }
\label{flc5}
\end{figure}
\begin{figure}[htb]
\begin{center}
\mbox{\epsfig{file=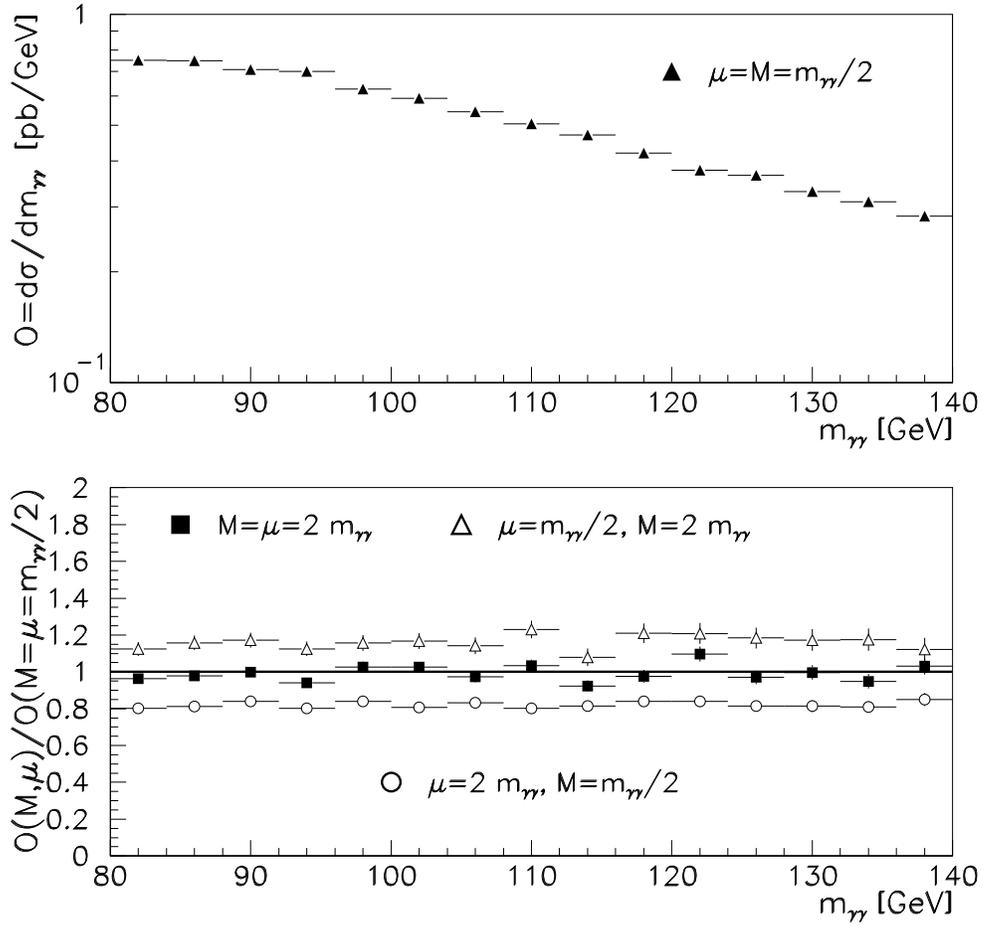,height=14cm}}
\end{center}
\caption{
 Diphoton differential cross section $d\sigma/dm_{\gamma \gamma}$ vs. 
 $m_{\gamma \gamma}$, the invariant mass of photon pairs, at LHC,
 $\sqrt{S}=14$ TeV without isolation. The following kinematic cuts are applied:
 $p_{T}(\gamma_{1}) \geq 40$ GeV, $p_{T}(\gamma_{2}) \geq 25$ GeV, 
 $|y(\gamma_{1,2})| \leq 2.5$. The scale dependence is shown on the bottom plot.
 $M = M_{f}$ is understood.
 }
\label{fig.scale.dep.incl}
\end{figure}
\begin{figure}[htb]
\begin{center}
\mbox{\epsfig{file=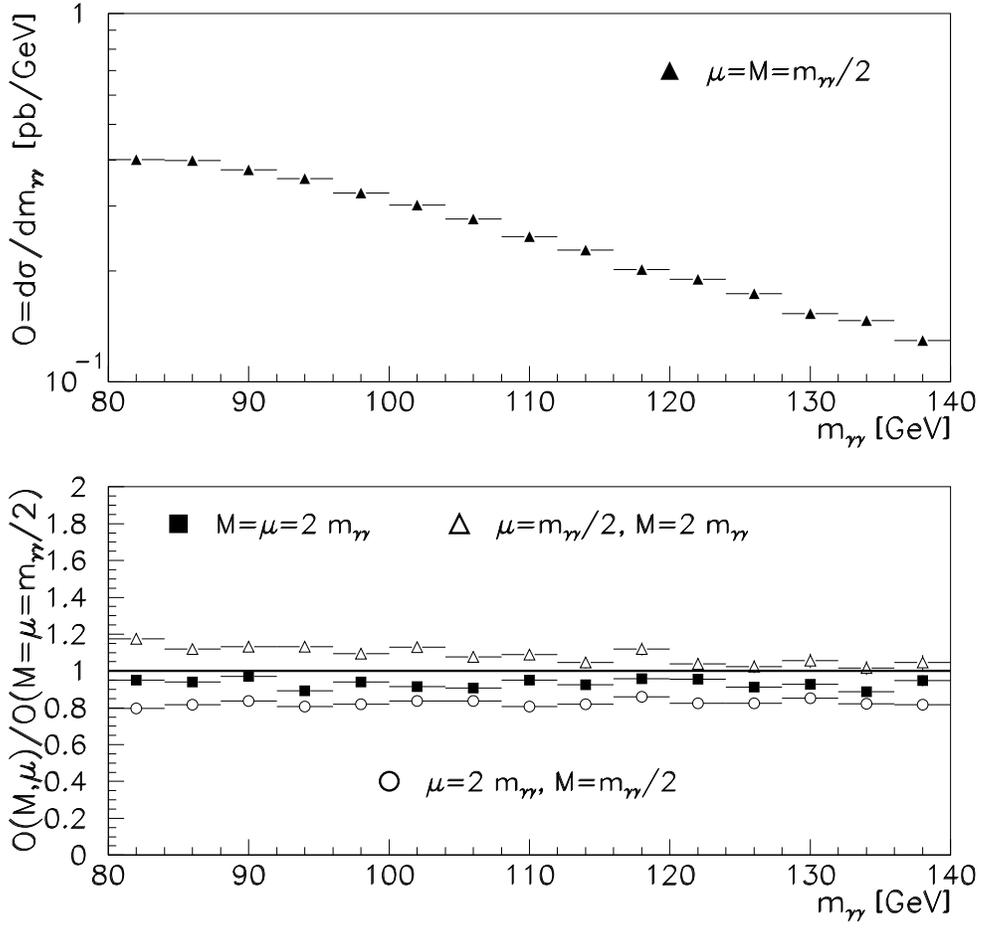,height=14cm}}
\end{center}
\caption{
 Diphoton differential cross section $d\sigma/d m_{\gamma \gamma}$ vs.
 $m_{\gamma \gamma}$ at LHC, $\sqrt{S}=14$ TeV, with isolation criterion
 $E_{T max}=5$ GeV in $R=0.4$. Same kinematic cuts as in fig. 
 \ref{fig.scale.dep.incl}. The scale dependence is shown on the bottom plot.
 $M = M_{f}$ is understood.
 }
\label{fig.scale.dep.isol}
\end{figure}
\begin{figure}[htb]
\begin{center}
\mbox{\epsfig{file=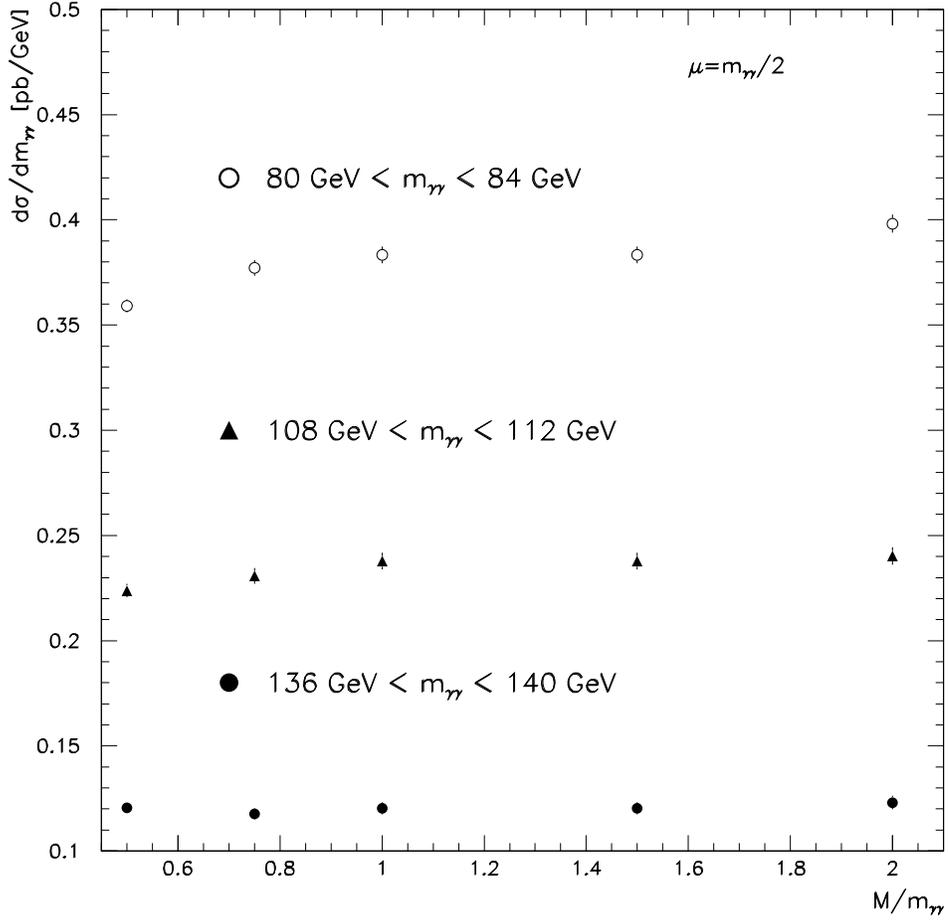,height=14cm}}
\end{center}
\caption{
 $M$ dependence of the ``direct+box" contribution to 
 $d\sigma/d m_{\gamma \gamma}$ in several $m_{\gamma \gamma}$
 bins at LHC, $\sqrt{S}=14$ TeV, with isolation criterion
 $E_{T max}=5$ GeV in $R=0.4$. Same kinematic cuts as in fig. 
 \ref{fig.scale.dep.incl}. $\mu$ is chosen to be $m_{\gamma \gamma}/2$, while 
 $M$ is varied between $m_{\gamma \gamma}/2$ and $2 m_{\gamma \gamma}$.
 }
\label{Mmonotonous}
\end{figure}
\begin{figure}[htb]
\begin{center}
\mbox{\epsfig{file=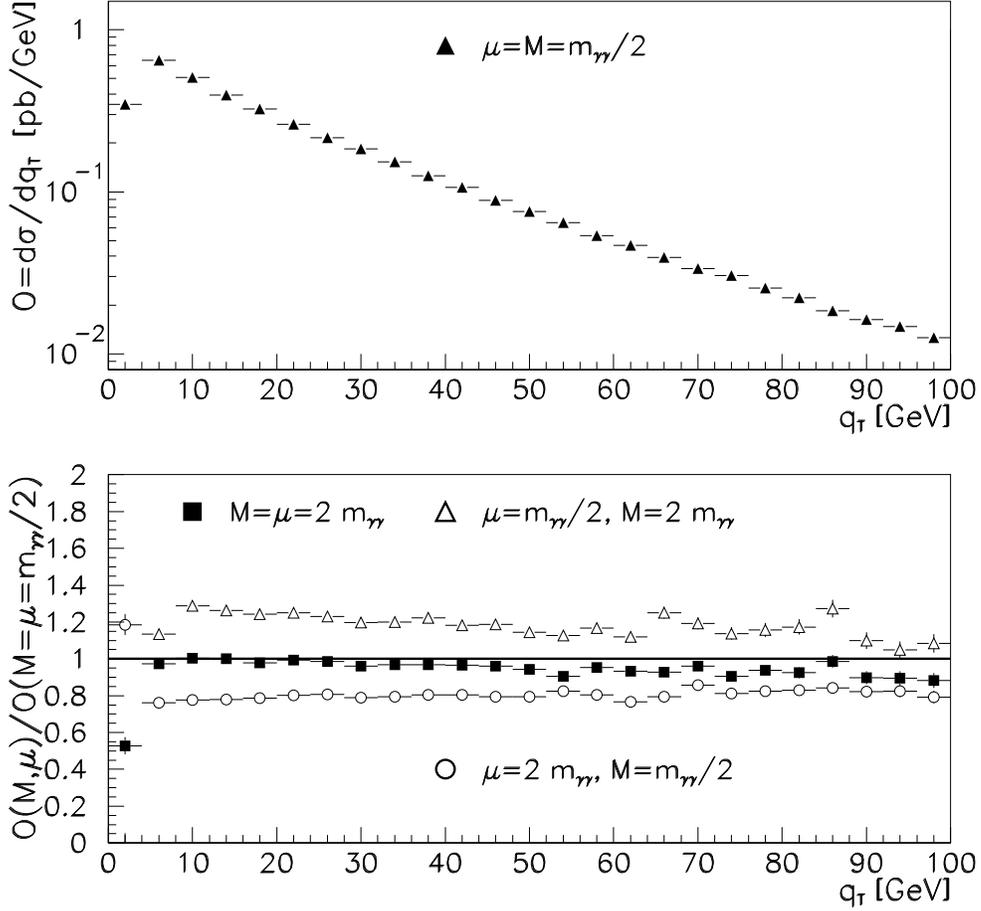,height=14cm}}
\end{center}
\caption{
 Diphoton differential cross section $d\sigma/dq_T$
 at LHC, $\sqrt{S}=14$ TeV, with isolation criterion
 $E_{T max}=5$ GeV in $R=0.4$. The following kinematic cuts are applied:
 $p_{T}(\gamma_{1}) \geq 40$ GeV, $p_{T}(\gamma_{2}) \geq 25$ GeV, 
 $|y(\gamma_{1,2})| \leq 2.5$, and 80 GeV $ \leq m_{\gamma \gamma} \leq $ 140 GeV.
 The scale dependence is shown on the bottom plot. $M=M_f$ is understood.
 }
\label{qt_lhc_5GeV}
\end{figure}
\begin{figure}[htb]
\begin{center}
\mbox{\epsfig{file=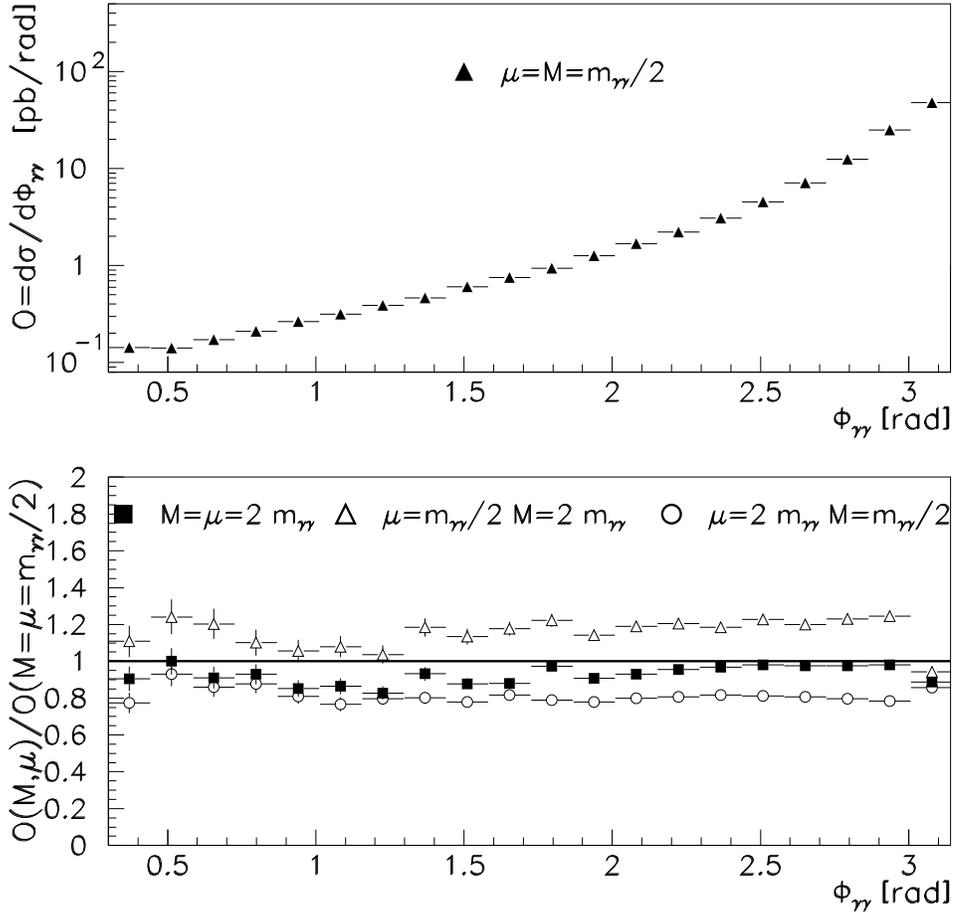,height=14cm}}
\end{center}
\caption{
 Diphoton differential cross section $d\sigma/d \phi_{\gamma \gamma}$ vs. 
 $\phi_{\gamma \gamma}$, the azimuthal angle between the two photons,
 at LHC, $\sqrt{S}=14$ TeV, with isolation criterion $E_{T max}=15$ GeV 
 in $R=0.4$.
 Same kinematic cuts as in fig. \ref{qt_lhc_5GeV}.
 The scale dependence is shown on the bottom plot. $M=M_f$ is understood.
  }
\label{phi_lhc_5GeV}
\end{figure}
\begin{figure}[htb]
\begin{center}
\mbox{\epsfig{file=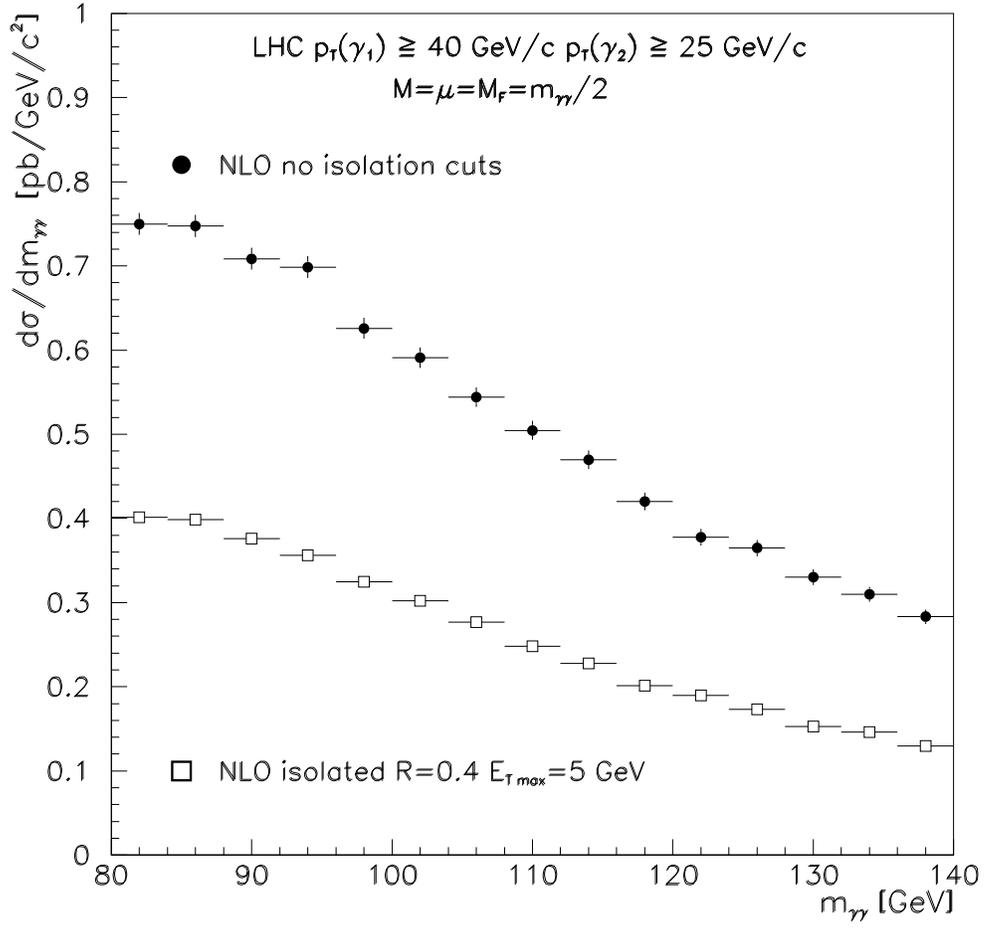,height=14cm}}
\end{center}
\caption{
 Diphoton differential cross section $d\sigma/d m_{\gamma \gamma}$ vs.
 $m_{\gamma \gamma}$ at LHC, $\sqrt{S}=14$ TeV, without and 
 with isolation criterion $E_{T max}=5$ GeV in $R=0.4$. 
 Same kinematic cuts as in fig. 
 \ref{fig.scale.dep.incl}. 
 The scale choice is $M=M_f=\mu=m_{\gamma \gamma}/2$.
 }
\label{incliso}
\end{figure}
\begin{figure}[htb]
\begin{center}
\mbox{\epsfig{file=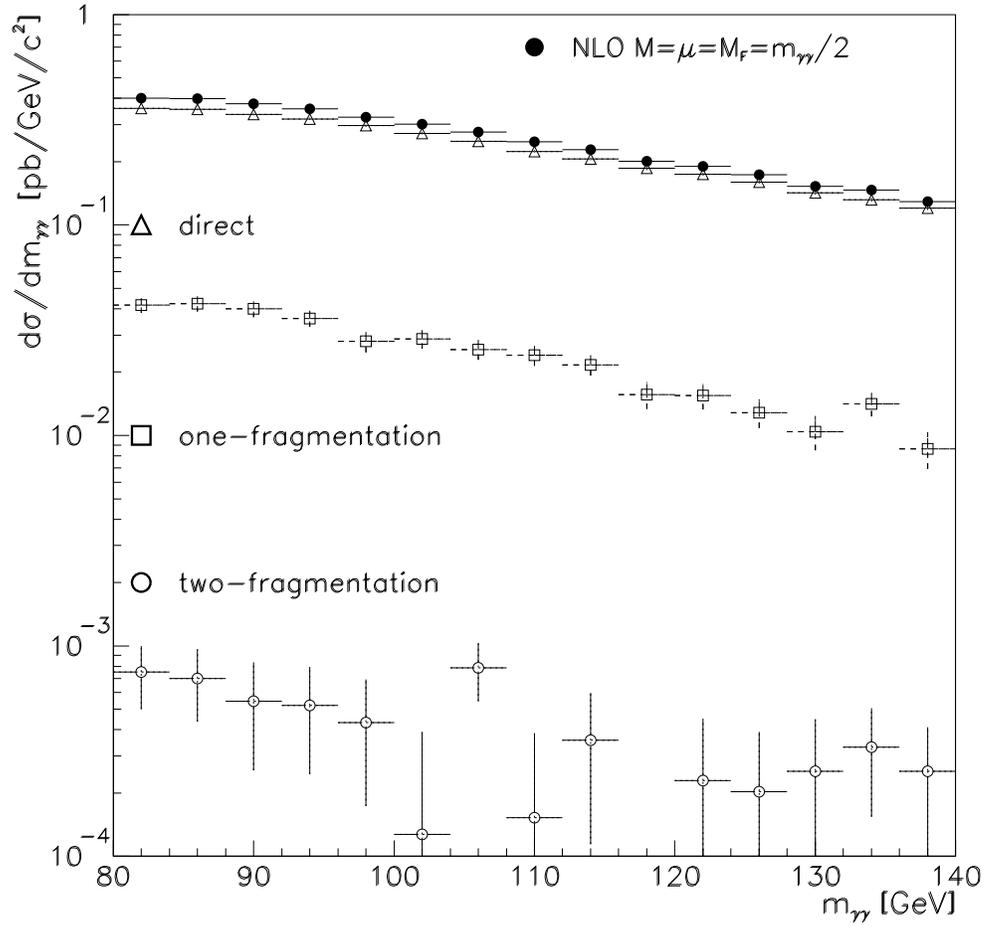,height=14cm}}
\end{center}
\caption{
 Splitting of the diphoton differential cross section 
 $d\sigma/dm_{\gamma \gamma}$ at LHC,
 $\sqrt{S}=14$ TeV with isolation criterion $E_{T max}=5$ GeV in $R=0.4$, 
 into the ``direct",``one
 fragmentation" and ``two fragmentation" components, shown for the scale 
 choice $\mu = M = M_{f}=m_{\gamma\gamma}/2$. 
 The following kinematic cuts are applied:
 $p_{T}(\gamma_{1}) \geq 40$ GeV, $p_{T}(\gamma_{2}) \geq 25$ GeV, 
 $|y(\gamma_{1,2})| \leq 2.5$. 
 }
\label{flc55}
\end{figure}
 \begin{figure}[htb]
\begin{center}
\mbox{\epsfig{file=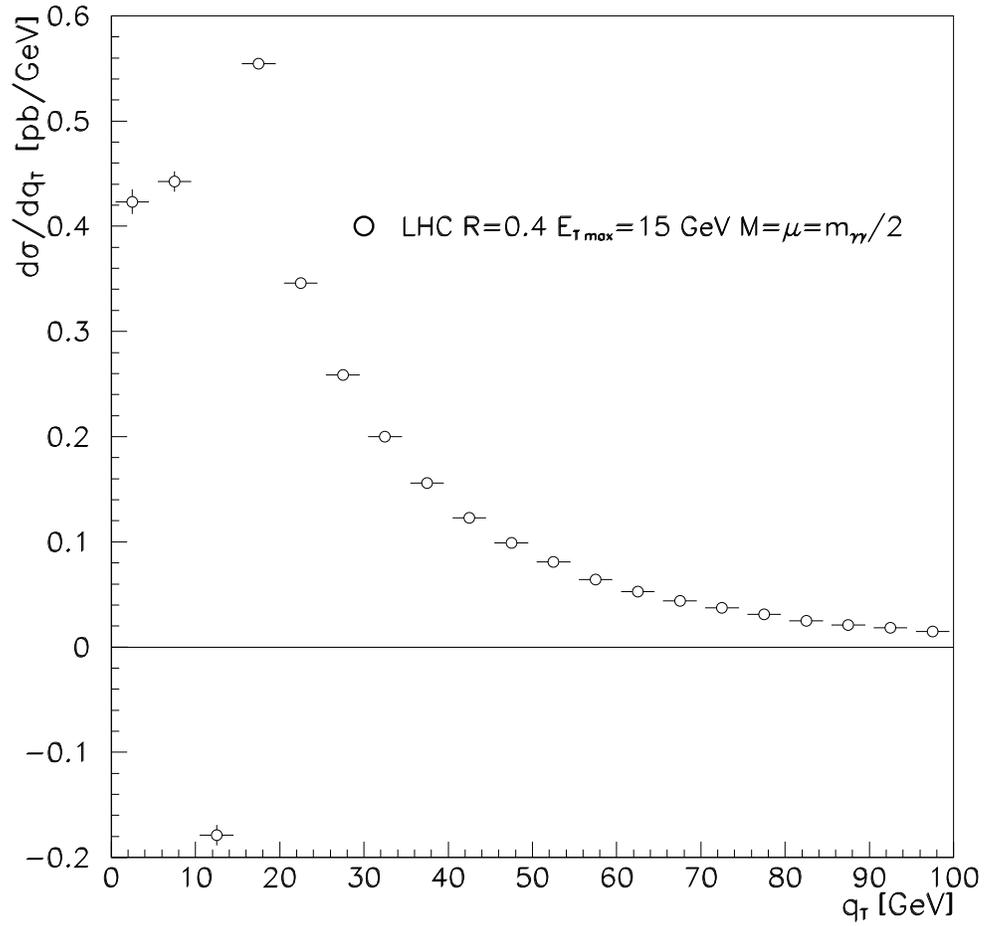,height=14cm}}
\end{center}
\caption{
 Diphoton differential cross section $d\sigma/dq_T$ at LHC, $\sqrt{S}=14$ TeV,
 with isolation criterion 
 $E_{T max}=15$ GeV in $R=0.4$. Same kinematic cuts as in fig. 
 \ref{fig.scale.dep.incl}.
 }
\label{flc13}
\end{figure}
\begin{figure}[htb]
\begin{center}
\mbox{\epsfig{file=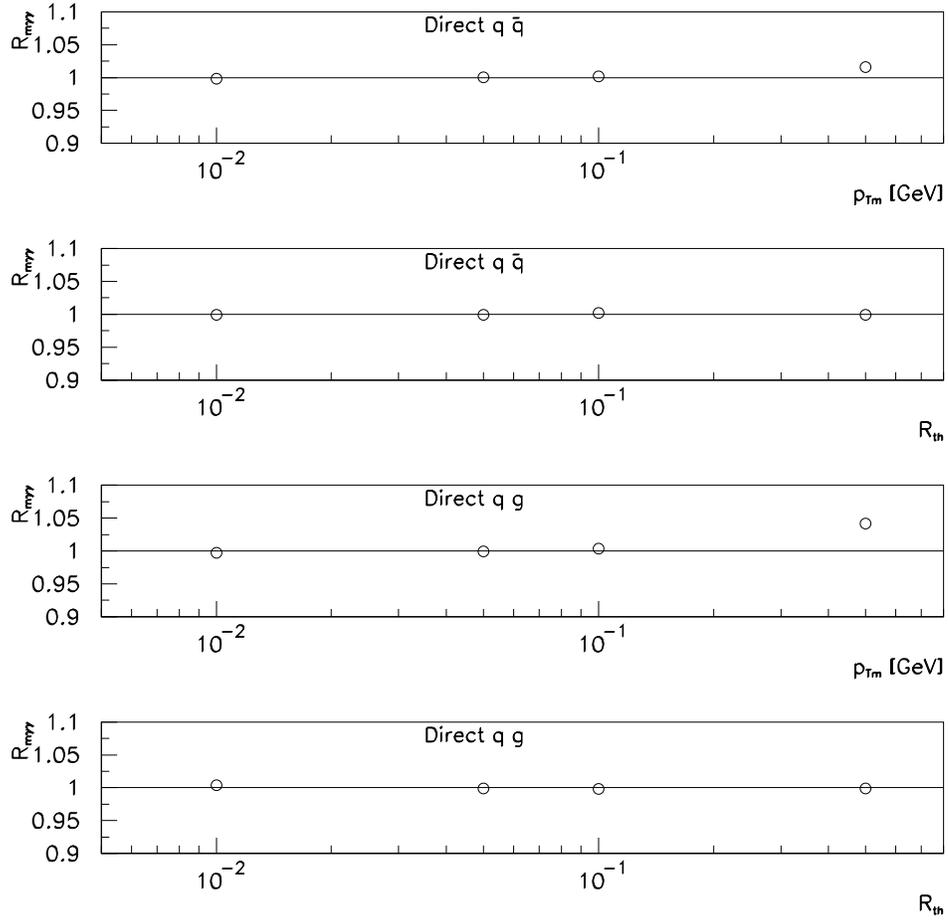,height=14cm}}
\end{center}
\caption{
Dependence of the ratio $R_{m_{\gamma \gamma}}$ (see equation~(\protect\ref{eqDEFRMGG})) over the phase space slicing parameters $R_{th}$ and $p_{Tm}$ for the ``direct" contribution.
}
\label{ptm_dep_dir}
\end{figure}

\begin{figure}[htb]
\begin{center}
\mbox{\epsfig{file=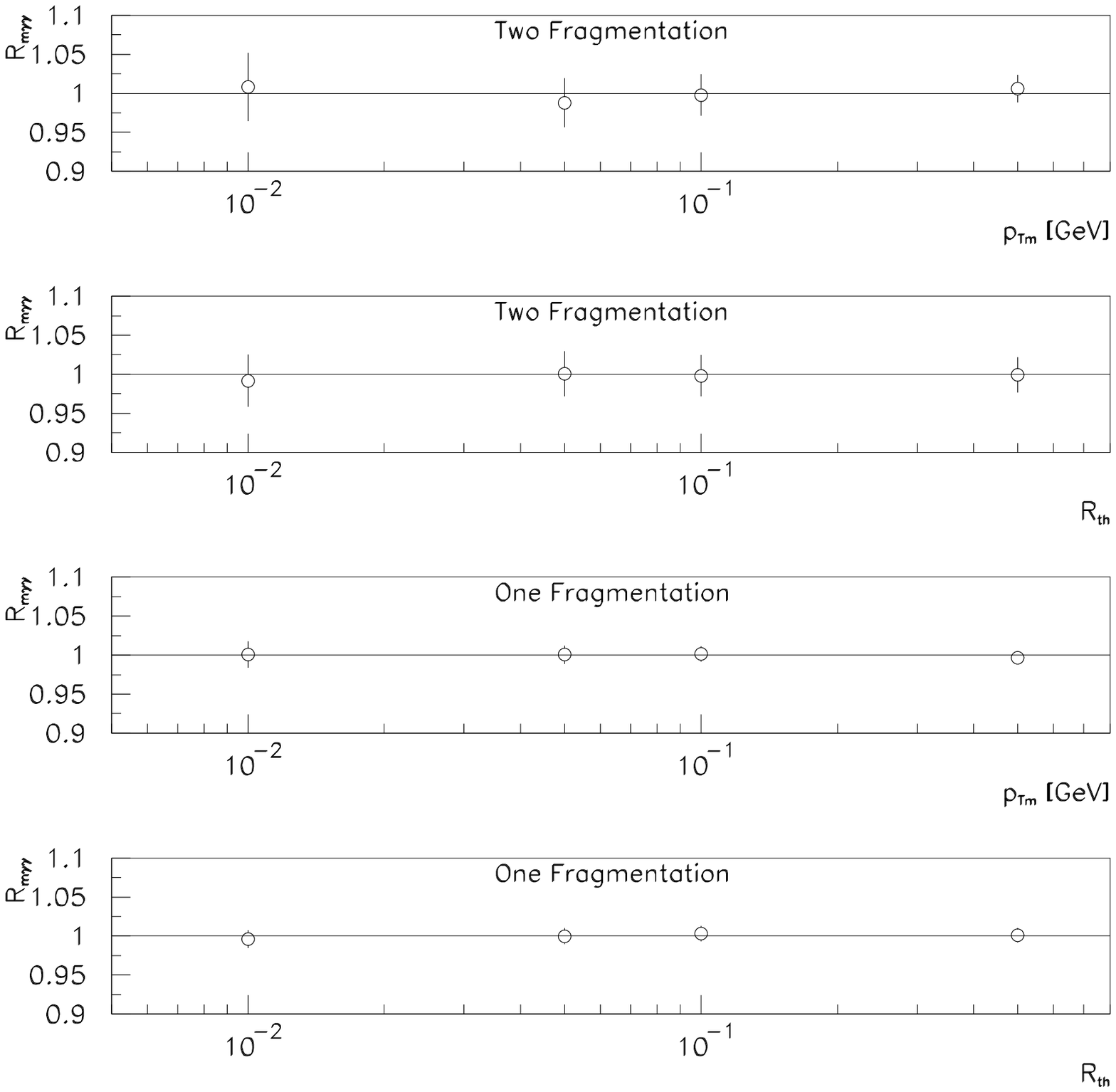,height=14cm}}
\end{center}
\caption{
Dependence of the ratio $R_{m_{\gamma \gamma}}$ (see equation~(\protect\ref{eqDEFRMGG})) over the phase space slicing parameters $R_{th}$ and $p_{Tm}$ for the ``one-" and ``two fragmentation" contributions.
}
\label{ptm_dep_frag}
\end{figure}


\begin{thebibliography}{99}

\bibitem{wa70cross}
WA70 Collaboration, E.~Bonvin {\em et al.},
 Z. Phys. {\bf C41} (1989) 591.

\bibitem{wa70correl}
WA70 Collaboration, E.~Bonvin {\em et al.},
Phys. Lett. {\bf 236B} (1990) 523.

\bibitem{e706}
Private communication from M.~Begel (E706 collaboration).

\bibitem{ua2}
UA2 Collaboration. J. Alitti {\em et al.},
Phys. Lett. {\bf 288B} (1992) 386.

\bibitem{cdf1}
CDF Collaboration, F. Abe {\em et al.},
Phys. Rev. Lett. {\bf 70} (1993) 2232.

\bibitem{d0}
Wei Chen, PhD Thesis (Univ. New-York at Stony Brook), Dec. 1997, unpublished; \\
D0 Collaboration (P. Hanlet for the collaboration), 
Nucl. Phys. Proc. Suppl. {\bf 64} (1998) 78.\\
All the numbers used in this article are taken from tables in Wei Chen PhD 
Thesis.
 
\bibitem{abfs}
P.~Aurenche, R.~Baier, A. Douiri, M.~Fontannaz and D.~Schiff,
Z. Phys. {\bf C29} (1985) 459; \\
P.~Aurenche, M.~Bonesini, L.~Camilleri, M.~Fontannaz and M.~Werlen,
Proceedings of LHC Aachen Workshop CERN-90-10 
G.~Jarlskog and D.~Rein eds., vol. II, p. 83.

\bibitem{owens}
B.~Bailey, J.~Ohnemus and J.F.~Owens,
Phys. Rev. {\bf D46} (1992) 2018;\\
B.~Bailey and J.F.~Owens,
Phys. Rev. {\bf  D47} (1993) 2735; \\
B.~Bailey and D.~Graudenz,
Phys. Rev. {\bf D49} (1994) 1486.

\bibitem{yuan}
C.~Balazs, E.L.~Berger, S.~Mrenna and C.P.~Yuan, 
Phys. Rev. {\bf D57} (1998) 6934; \\
C.~Balazs and C.P.~Yuan, 
Phys. Rev. {\bf D59} (1999) 114007.

\bibitem{phofrag}
L.~Bourhis, M.~Fontannaz and J.Ph.~Guillet,
Eur. Phys. J. {\bf C2} (1998) 529.
  
\bibitem{dgsz}
A. Djouadi, D. Graudenz, M. Spira and P. Zerwas,
Nucl. Phys. {\bf B453} (1995) 17. 

\bibitem{delduca}
V.~Del Duca, W.B.~Kilgore and F.~ Maltoni, hep-ph/9910253. 

\bibitem{deflorian}
D.~de Florian and Z.~Kunszt, Phys. Lett. {\bf 460B} (1999) 184;\\
C.~Balazs, P.~Nadolsky, C.~Schmidt and C.P.~Yuan, hep-ph/9905551.

\bibitem{smirnov-tausk}
V.A.~Smirnov, Phys. Lett. {\bf 460B} (1999) 397;\\
J.B.~Tausk, hep-ph/9909506.
 
\bibitem{cfg}  
P.~Chiappetta, R.~Fergani and J.-Ph~Guillet, 
Z. Phys. {\bf C69} (1996) 443.

\bibitem{slicing}
M.A.~Furman \NP {B197} (1982) 413; \\
W.T.~Giele and E.W.N.~Glover,  Phys. Rev. {\bf D46} (1992) 1980;\\
W.T.~Giele, E.W.N.~Glover and D. Kosower, \NP{B403} (1993) 633.

\bibitem{subtraction}
R.K.~Ellis, D.A.~Ross and A.E.~Terrano, \NP {B187} (1981) 421; \\
S.~Frixione, Z.~Kunszt and A.~Signer, Nucl. Phys. {\bf B467} (1996) 399;\\
S.~Catani and M.H.~Seymour, \NP {B485} (1997) 291.

\bibitem{bases}
S. Kawabata, Comp. Phys. Comm. {\bf 88} (1995) 309.

\bibitem{paw}
R.~Brun, O.~Couet, G.E.~Vandoni and P.~Zabarini,
Comp. Phys. Comm {\bf 57} (1989) 432; \\ 
PAW, Physics Analysis Workstation, CERN Program Library Q121 \\
(http://wwwinfo.cern.ch/asd/paw/index.html).
   
\bibitem{pythia}
H.-U.~Bengtsson and T.~Sj\"ostrand, Comp. Phys. Comm. {\bf 46} (1987) 43; \\
T.~Sj\"ostrand, Comp. Phys. Comm. {\bf 82} (1994) 74; \\
T. Sj\"ostrand, LU TP 95-20, hep-ph/9508391.

\bibitem{herwig}
G.~Marchesini and B.R.~Webber, \NP {B 310} (1988) 461; \\
G.~Marchesini, B.R.~Webber, G.~Abbiendi, 
I.G.~Knowles, M.H.~Seymour and L.~Stanco, hep-ph/9607393.

\bibitem{frixione}
S. Frixione, Phys. Lett. {\bf 429B} (1998) 369.

\bibitem{boo1}
H. Baer, J. Ohnemus and J.F. Owens, Phys. Rev. {\bf D42} (1990) 61.

\bibitem{abf}
P. Aurenche, R. Baier and M. Fontannaz, Phys. Rev. {\bf D42} (1990) 1440.

\bibitem{berger-qiu}
E.L. Berger and J. Qiu, Phys. Rev. {\bf D44} (1991) 2002.

\bibitem{gordon-vogelsang}
L.E. Gordon and W. Vogelsang, Phys. Rev. {\bf D50} (1994) 1901.

\bibitem{cfgp}
S. Catani, M. Fontannaz and E. Pilon, work in preparation.

\bibitem{Kunszt-Troscanyi}
Z. Kunszt and Z. Troscanyi, Nucl. Phys. {\bf B394} (1993) 139.

\bibitem{berger-guo-qiu}
E.L.~Berger, X.F.~Guo and J.W.~Qiu, 
Phys. Rev {\bf D54} (1996) 5470.


\bibitem{cfp}
S. Catani, M. Fontannaz and E. Pilon, Phys. Rev. {\bf D58} (1198) 094025.

\bibitem{glover-morgan}
E.W.N. Glover and A.G. Morgan, Z. Phys. {\bf C62} (1994) 311.

\bibitem{abfow}
P.~Aurenche, R.~Baier, M.~Fontannaz, J. Owens and M. Werlen,
Phys. Rev. {\bf D39} (1989) 3275.

\bibitem{abfkw}
P.~Aurenche, R.~Baier, M.~Fontannaz, M.N.~Kienzle-Foccacci and M. Werlen,
Phys. Lett. {\bf 233B} (1989) 517.

\bibitem{smrs}
P.J.~Sutton, A.D.~Martin, R.G. Roberts and W.J.~Stirling,
Phys. Rev. {\bf D45} (1992) 2349.

\bibitem{grv}
M.~Gl\"uck, E.~Reya and A.~Vogt,
Z. Phys. {\bf C53} (1992) 651.


\bibitem{Wmass-d0}
B. Abbott {\it et al.}, Phys. Rev. {\bf D58} (1998) 092003

\bibitem{mrs99}
A.D.~Martin, R.G.~Roberts, W.J.~Stirling and R.S.~Thorne, hep-ph/9907231.

\bibitem{wormesley_privatecom}
J. Womersley, private communication.

\bibitem{owens_privatecom}
J.F. Owens, private discussion.

\bibitem{cms}
CMS technical proposal CERN/LHCC 94-38 (p 179); \\
ATLAS technical proposal CERN/LHCC 94-43; \\
V.~Tisserand for the ATLAS collaboration,
in Proc. 6th Int. Conf. on Calorimetry in High Energy Physics (ICCHEP 96),
ed. by A.~Antonelli, S.~Bianco, A.~Calcaterra, F.L.~Fabbri,
(Frascati physics series; 6) p 475.

\bibitem{reviewir}
Y.~Dokshitzer, D.~Dyakonov and S.~Troyan, Phys. Rep. {\bf 58} (1980) 269; \\
A.~Basseto, M.~Ciafaloni and G.~Marchesini, Phys. Rep. {\bf 100} (1983) 201 and
references therein;\\
see also \cite{cs}.

\bibitem{balacz}
C.~Balazs, private communication.
 
\bibitem{cs}
J.~Collins and D.~Soper, Nucl. Phys. {\bf B193} (1981) 381; \\
J.~Collins and D.~Soper, Nucl. Phys. {\bf B197} (1982) 446; \\ 
J. Collins, D. Soper and G. Sterman, Nucl. Phys. {\bf B250} (1985) 199.

\bibitem{sterman-kidonakis}
R.~Bonciani, S.~Catani, M.~Mangano and P.~Nason, Nucl. Phys. {\bf B529} (1998) 424; \\
N.~Kidonakis, G.~Oderda and G.~Sterman, Nucl. Phys. {\bf B531} (1998) 365.

\bibitem{catani-webber}
S.~Catani and B.~Webber, JHEP 9710:005 (1997). 

\bibitem{ellis-sexton}
R. K.~Ellis and J. C.~Sexton, \NP {B269} (1986) 445.
\end{thebibliography}
\end{document}